\documentclass[aps,prd,nofootinbib,floatfix,amsmath,amssymb]{revtex4}

\usepackage{graphicx}
\usepackage{dcolumn}
\usepackage{bm}
\usepackage{float}
\usepackage{color}
\usepackage{subfigure}
\usepackage{hyperref}
\usepackage[usenames, dvipsnames]{pstricks}
\usepackage{footnote} 

\def \dL{\mathcal{L}}

\newcommand{\be}{\begin{eqnarray}}
\newcommand{\ee}{\end{eqnarray}}
\newcommand{\nn}{\nonumber}
\newcommand{\yd}{{{Y^d_1}}}

\newcommand{\yyuu}{{{Y^u_2}}}

\newcommand{\hggs}{\Phi_1}
\newcommand{\hggss}{\Phi_2}
\newcommand{\hgt}{\tilde{\Phi}_1}
\newcommand{\hgtt}{\tilde{\Phi}_2}
\begin{document}
\preprint{HEP-BUAP-01.13}

\title{Flavor Violating Higgs signals in the Texturized Two-Higgs Doublet Model (THDM-Tx)}

\author{M. A. Arroyo-Ure\~na, J. Lorenzo Diaz-Cruz, Enrique D\'iaz, Javier A. Orduz-Ducuara}
\affiliation{%
Facultad de Ciencias F\'isico-Matem\'aticas\\
Benem\'erita Universidad Aut\'onoma de Puebla, C.P. 72570, Puebla, Pue., Mexico.}%

\date{\today}

\begin{abstract}
Flavor violating Higgs signals, such as the top FCNC decay $t\to ch^0$ and the LFV Higgs decay $h^0\to \tau\mu$, 
have been studied at the LHC.  These signals can arise within the general Two-Higgs doublet model (THDM), 
where each Higgs doublet couples to all fermions types through Yukawa matrices $Y^f_1$ and $Y^f_2$.  
The Yukawa matrices can be assumed to have the same form or they could have different 
structures. In this paper we study the case when both $Y^f_1$ and $Y^f_2$ have
completely different forms, but in such a way that they complement to produce 
a specific hermitian mass matrix. We find that for specific four-zero textures, 
the Flavor Violating Higgs couplings depend only on the free parameters $\tan \beta$, $\gamma_f$ and the 
fermion masses. We use the current bounds on the low energy processes, 
to derive constraints on the Heavy Higgs boson mass, $\tan \beta$ and $\gamma_f$.
 Then, we use these constraints to evaluate the LFV  Higgs decays, 
which reach Branching ratios that could be tested at the LHC.
\end{abstract}

\pacs{Valid PACS appear here}
\maketitle

\section{\label{sec:level1}Introduction}

The discovery at the Large Hadron Collider (LHC) of a new particle with 
Standard Model (SM)-Higgs like  properties, and mass $M_h=125.09 \pm 0.21 (stat.) \pm 0.11 (syst.)$ GeV \cite{Higgs:higgs-atlas-cms}, 
seems to confirm the linear realization of the mechanism of Electroweak symmetry breaking. 
This is needed in order to induce the masses of gauge bosons and fermions within the SM \cite{Gunion:1989we}. 
Furthermore, the current experimental studies at LHC are testing the Higgs 
properties at levels that could allow to discriminate between the minimal SM Higgs doublet 
and other extensions of the SM that include more complicated Higgs sectors \cite{DiazCruz:2003qs}.

In our approach we consider a possible link between the Higgs sector and flavor physics that could be revealed
by a thorough study of the THDM with textures (THDM-Tx). In any case such models have a rich phenomenology, 
with lots of interesting signals that could be searched at future colliders.
One of the simplest proposals for physics Beyond the Standard Model, is the
so called Two-Higgs Doublet Model (THDM),  which was initially studied in connection with
the search for the origin of CP violation \cite{Deshpande:1977rw}, and later it was used 
in connection with other theoretical ideas in particle physics, 
such as supersymmetry \cite{Martin:1997ns}, extra dimensions \cite{Quiros:2006my} 
and strongly interacting systems \cite{Pomarol:2012sb}, \cite{Aranda:2007tg}.

Several possible realizations of the general THDM have been considered in the
literature, which have come to be known as Type I, II and III. There
are also other models called X, Y, Z, but in some sense they can be considered variations of the above
models (for a review see \cite{Branco:2011iw}). Model I can have an exact discrete symmetry $Z_2$, which permits a possible dark
matter candidate coming from the $Z_2-$odd scalar doublet \cite{Ginzburg:2010wa}. Within type I models, a single 
Higgs doublet gives mass to the up, down quarks and
leptons. The type II model \cite{Ginzburg:2004vp} assigns one doublet to each fermion type, then according 
to the Glashow-Weinberg Theorem \cite{Glashow:1976nt}, this 
suffices to avoid Flavor Changing Neutral Currents (FCNC) mediated by the Higgs bosons; 
this type II model also arises in the minimal SUSY extension of the SM \cite{Gunion:1988yc}.

In the most general version of the THDM both Higgs 
doublets couple to all types of fermions. 
In this case, the diagonalization of the full mass  matrix, does not imply that each Yukawa matrix 
is diagonalized, therefore FCNC can appear at tree level. 
Within this general model, one must reproduce the 
observed fermion masses and mixing angles, while at the same time the level of  FCNC 
must satisfy current experimental bounds \cite{Fritzsch:1995nx, Branco:1999nb, Hall:1993ca}. 
One possibility to achieve this, is the assumption that the Yukawa matrices 
have a certain texture form, i.e. with zeroes in different elements.

The general model has been previously referred to as the THDM of type III. However, this naming scheme 
has become confusing, in part because it has also been used to denote a different type of model \cite{Altmannshofer:2012ar}, 
but also because some specific cases have acquired a relevance of their own. Among the relevant sub-cases 
of the general THDM, one can include the so-called Minimal Flavor Violating (MFV) THDM \cite{Buras:2010mh}, 
which is thought to provide precisely the minimal level of FCNC consistent with data. MFV models could be studied 
from a pure phenomenological point of view \cite{D'Ambrosio:2002ex}, or as arising from flavor 
symmetries \cite{Aranda:2012bv}. Although the so-called THDM with Alignment does not contain 
flavor violation, it is another possibility one can use to obtain realistic models \cite{Pich:2009sp}. 
Thus, in order to clarify the notation and to single out the use of textures within the THDM,
from now on we shall call the two-Higgs doublet model with textures as THDM-Tx.

One of the first studies of the THDM with textures \cite{Cheng:1987rs} considered a specific form with six-zeroes, 
as well as other variations with cyclic textures. In that work it was 
identified that the texture assumption implies a specific pattern of FCNC Higgs-fermion couplings,  
known nowadays as  the Cheng-Sher ansatz, which is of size $\frac{\sqrt{m_i m_j }}  {v}$. It was found that such vertex could satisfy FCNC bounds with 
Higgs masses lighter than  O(TeV). The extension of the THDM-Tx with a four-zero texture was presented 
in \cite{Zhou:2003kd, DiazCruz:2004tr}. 
The phenomenological consequences of these textures (Hermitian 4-textures or non-hermitian  6-textures) were 
considered  in \cite{DiazCruz:2004pj}, while further phenomenological studies were presented in
\cite{Wu:2004kr, Li:2005rr, Carcamo:2006dp}.

Several models for Yukawa matrices could be identified which lead to specific patterns of 
flavor violating Higgs interactions. For instance, one can assume that the Yukawa matrices $Y^f_1$ and $Y^f_2$
have the same form (a case that we call ``Parallel Textures"). It is also possible to have a
Yukawa matrix, say $Y^f_1$, with some specific texture, while the second matrix $Y^f_2$ has only some elements 
different from zero, at positions that coincide with some elements of $Y^f_1$, as in the so called top-specific models 
discussed in the literature \cite{Atwood:2005bf}, we call this case ``Semi-parallel textures".

In this paper we study another possibility, namely that $Y^f_1$ and $Y^f_2$ have completely
different structure. Namely, for specific four-zero textures that reproduce all fermion masses and the CKM matrix, 
these vertices only depend on the parameters $\tan \beta(=\frac{v_2}{v_1})$ and $\gamma_f$ $(0 < \gamma_f <1, f=u,d,l)$, which  
appears in the relation between the third family mass and the 33 entry of the corresponding mass matrix.
 We use the current bounds on the low energy processes, to impose constraints on the values 
of $\gamma_f$ and $\tan \beta$.
We also compare these constraints with the ones obtained for the cases of parallel and fermion-specific
textures.
Furthermore, we also include the constraints obtained from current LHC bounds on the Higgs boson couplings, and derive 
predictions for the Flavor Violating decays $h^0\to\tau\mu$ and $t\to ch^0$,  which reach Branching ratios 
that could be tested at the LHC in the forthcoming era of precision Flavor Higgs Physics.

The organization of our paper is as follows. A classification of the different types of 
Yukawa matrices that produce a mass matrix with four-zero textures, 
as well as the diagonalization  of the mass matrix, is presented in section \ref{sec:class-textu}.
Section \ref{sec:level2} discusses generalities of the THDM-Tx,
including the Yukawa interaction Lagrangian written in terms of mass eigenstates.
Low energy constraints are discussed in section \ref{sec:cons-low-ener}, 
including $K-\bar{K}$ and $B-\bar{B}$ mixing, 
$B_s^0 \to \mu^+\mu^-$, $\mu\rightarrow e \gamma$, $\tau^-\rightarrow \mu^-\mu^+\mu^-$, $B\to D(D^*)\tau\nu$ and $\Delta a_{\mu}$. 
Constraints from current Higgs searches at LHC are included in section \ref{sec:LHC-signal}. 
The prediction of our model for the decays $h^0 \to \tau \mu$ and $t \to c h^0,$ are discussed in 
section \ref{sec:predictions}. Conclusions of our work are presented 
in section \ref{sec:conclusio}, 

\section{Fermion Mass Matrix: Diagonalization and  Texture Patterns  }\label{sec:class-textu}

Within the general THDM, each Higgs doublet couples to fermions of type $f$ ($f=u,d,l$) through the
Yukawa matrices $Y^f_1$ and $Y^f_2$. After Spontaneous Symmetry Breaking (SSB), these matrices combine to produce a fermion mass matrix with some structure. The mass matrix for each fermion type $f(=u,d,l)$  receives contributions from both vevs $v_1$ and $v_2$, i.e.

\begin{equation}
{M_f}=\frac{1}{\sqrt{2}} \left(v_1 {Y}^f_1 + v_2 {Y}^f_2 \right), 
\end{equation}

To obtain physical fermion masses we need to diagonalize the mass matrix; this is
achieved through a bi-unitary transformation $\mathcal{O}_f=V_f^{\dagger} P_f$, i.e.

\begin{equation}
{M_D}=\mathcal{O}_f{M_f}\mathcal{O}_f^\dagger = \mathcal{O}_f \frac{1}{\sqrt{2}} 
\left(v_1 {Y}^f_1 + v_2 {Y}^f_2 \right)\mathcal{O}_f^\dagger,
\end{equation}
the form of the matrix $\mathcal{O}_f$ depends on the texture type; closed forms
have been obtained for the 4- and 6-texture hermitian and non-hermitian cases.
Although $\mathcal{O}_f$ diagonalizes the matrix $M_f$, it does not necessarily diagonalize each of the 
Yukawa matrices that make up $M_f$, thus neutral flavor violating Higgs-fermion interactions will be induced.

\subsection{Diagonalization of the Fermion mass matrices of  four-texture type}
 For the 4-texture case the mass matrix takes the form:

\begin{eqnarray} 
{M_f} =  \left( \begin{array}{ccc}
0 & D & 0 \\
D^* & C & B \\
0 & B^* & A
\end{array} \right),\label{eq:mass}
\end{eqnarray} 
then $\mathcal{O}_f$ is given by:

\begin{eqnarray}
\small \mathcal{O}_f =
\left( \begin{array}{ccc}
\sqrt{\frac{m_2 m_3 (A -m_1 )}{A  (m_2-m_1 ) (m_3-m_1 )}}
& \sqrt{\frac{m_1 m_3 (m_2-A )}{A (m2-m_1 ) (m_3-m_2 )}}
& \sqrt{\frac{m_1m_2 (A-m_3 )}{A (m_3-m_1 ) (m_3-m_2 )}} \\
-\sqrt{\frac{m_1 (m_1-A )}{ (m_2-m_1 ) (m_3-m_1 )}}
& \sqrt{\frac{m_2 (A-m_2 )}{ (m_2-m_1 ) (m_3-m_2 )}}
& \sqrt{\frac{m_3 (m_3-A )}{ (m_3-m_1 ) (m_3-m_2 )}} \\
\sqrt{\frac{m_1 (A-m_2 ) (A-m_3 )}{A (m_2-m_1 ) (m_3-m_1 )}}
& -\sqrt{\frac{m_2 (A-m_1 ) (m_3-A )}{A (m_2-m_1 ) (m_3-m_2 )}}
& \sqrt{\frac{m_3(A-m_1 ) (A-m_2 )}{A (m_3-m_1 ) (m_3-m_2 )}}
\end{array} \right),
\end{eqnarray} \\
and

\begin{eqnarray}\label{matrizP}
P_f =
\left( \begin{array}{ccc}
1 & 0               & 0\\
0 & e^{i\alpha_1} & 0\\
0 & 0               & e^{i\alpha_2} \\
\end{array} \right).
\end{eqnarray}
where $m_i$ $(i=1, 2, 3)$ are the fermion masses. Hence, following refs. \cite{Fritzsch:1995nx, Branco:1999nb}, we use  $det(M_f)= - D^2 A = m_{{1}}m_{{2}} m_{{3}}$
and assume the ordering $m_{{3}}>A>m_{{2}} >m_{{1}}$, so that 
$A - m_{{3}}<0$, to get real mixing angles we take $m_{{1}}<0$.

From these expressions we find a relation between the components of the 4-texture mass matrix and the 
physical fermion masses, which will be useful in order to find the expressions for the Higgs-fermion
interactions, namely:

\begin{eqnarray}
A &=& m_{{3}} \left(1-r_2 \gamma_f \right), \\
B &=& m_{{3}} \sqrt{\frac{r_{{2}} \gamma_f  \left(r_{{2}} \gamma_f +r_{{1}}-1\right) \left(r_{{2}} \gamma_f +r_{{2}} - 1\right)}{1-r_{{2}} \gamma_f}}, \\
C &=& m_{{3}} \left(r_2 \gamma_f +r_{{1}}+r_{{2}}\right), \\
D &=& \sqrt{-\frac{m_{{1}} m_{{2}}}{1-r_{{2}} \gamma_f },} 
\end{eqnarray}
\nolinebreak
where $r_{{i}} = \frac{m_{{i}}}{m_{{3}}}.$ 
Thus, the relation between the third family mass and the 33 entry of the mass matrix 
$A = m_{{3}} (1 - r_{{2}} \gamma_f) $, depends on the parameter 
$\gamma_f$ ($0 < \gamma_f < 1$), where $f=u,d,l$. 
Concerning the construction of the physical 
CKM matrix $V_{CKM}=\mathcal{O}_u\mathcal{O}_d^\dagger$, we are able to correctly reproduce the values of CKM matrix \cite{PDG:2016}; 
for example for the values of $\gamma_u=0.13$, $\gamma_d=0.1$, $\alpha_1^u=2.473555$, $\alpha_2^u=0.65$, $\alpha_1^d=1.045$ and $\alpha_2^d=1.69$, we obtain

\begin{eqnarray}
{V}_{CKM}^{THDM-Tx} =  \left( \begin{array}{ccc}
0.97424 & 0.22548 & 0.00294 \\
0.22530 & 0.97342 & 0.04100 \\
0.00918 & 0.04000 & 0.99915
        \end{array} \right).
\end{eqnarray}

\subsection{Classification of Textures: Parallel, Semi-Parallel and Complementary}

For the purpose of studying the fermion masses and CKM mixing matrix, 
it does not matter how each Yukawa matrix contributes to the mass matrix.
These matrices can be obtained from a variety of flavor symmetries, discrete or
continuos, local or global, in 4D or beyond. It could be interesting to look for 
some experimental signals that could help to discriminate among all those
different possibilities. It turns out that the flavor violating Higgs interactions
could provide such methodology. 

Rather than focusing on an specific model, we study here the different possibilities one
could use in order to arrive to some mass  matrix $M_f$ from the different patterns of Yukawa matrices
$Y^f_1$ and $Y^f_2$. In such case one can express one rotated Yukawa matrix in terms of the other one 
and the mass eigenvalues.
For instance we can rotate  $Y_1 $, and fix $Y_2$ through the relation:
$\tilde{Y_2} =\frac{\sqrt{2}}{v_2} \bar{M}_f - \cot\beta \tilde{Y}_1$.

{{I) ``Parallel textures".}} This is the most widely studied case, which assumes that both ${Y}^f_1$ 
and ${Y}^f_2$ have the same structure, namely: 

\be
Y_1 = 
\begin{pmatrix}
0 & d_1 & 0\\
d^*_1 & c_1 & b_1\\
0 & b^*_1 & a_1 \\
\end{pmatrix},
\hspace{5mm}
Y_2 = 
\begin{pmatrix}
0 & d_2 & 0\\
d^*_2 & c_2 & b_2\\
0 & b^*_2 & a_2 \\
\end{pmatrix}.
\ee

The explicit form of the 33 and 23 elements for the rotated Yukawa matrix $\tilde{Y}^f_2$
is given by:

\begin{eqnarray}
   ({\tilde{Y}}^f_2)_{23} &=& \frac{\sqrt{m_2m_3}}{v}\chi_{23},\\
 ({\tilde{Y}}^f_2)_{33} &=&\frac{m_3}{v}\chi_{33},
\end{eqnarray}
where
\begin{equation}
\chi_{23}  =  \sqrt{\frac{ m_{3} }{ m_{2} } } \left( \frac{b_1 v}{m_3 \tan \beta} -
                \frac{ F_1}{ \sin\beta } \right) e^{i\alpha_{2} } 
  +  \left( \frac{ (a_1-c_1)v }{m_3 \tan \beta }-\sqrt{2}\frac{ F_2}{ \sin\beta } \right) \sqrt{\gamma_{f}},
\end{equation}

\begin{equation}
\chi_{33}  = \sqrt{2}\frac{F_{1}}{Q}\left(\frac{b_1 v}{m_3\tan\beta}-\frac{F_{1}}{\sin\beta}\right)\cos\alpha_{2}
  - \frac{c_1v }{m_3\tan\beta}F_{3}+\sqrt{2}\frac{(Q-F_{3}R)}{\sin\beta}.
\end{equation}

We define $r_2=m_2/m_3$, $R=1- r_2 \gamma_f -r_2$,  $Q=1-r_2$, $G=r_2\gamma_f$, $F_1= \sqrt{2GR}$, $F_2= Q-2G$, $F_3=G/Q$. 
In this case, the elements $\chi_{23,33}$ depend on the complete set of parameters $a_1$, $b_1$, $c_1$, $\alpha_2$, $\gamma_f$ and $\beta$. 
In the literature \cite{DiazCruz:2004pj}, it is usually assumed that these elements already have 
the Cheng-Sher form, with coefficients $\chi^f_{ij}$ of $O(1)$, which are constrained from analyzing
the FCNC and LFV processes. 

{{II) ``Semi-Parallel textures".}} However, it is also possible that ${Y}^f_1$ and ${Y}^f_2$ could have different textures, 
but in such a way that the resulting mass matrix has a realistic 
texture. It could be that $Y^f_1$ has a certain texture, but $Y^f_2$ has some entry different
from zero in one of the entries that is also non-zero for $Y^f_1$. One example has $Y^f_1$
with a four-zero texture, while ${Y}^f_2$ has only a non-zero 33 entry, namely:
\begin{equation}
Y_1 = 
\begin{pmatrix}
0 & d_1 & 0\\
d^*_1 & c_1 & b_1\\
0 & b^*_1 & a_1 \\
\end{pmatrix},
\hspace{5mm}
Y_2 = 
\begin{pmatrix}
0 & 0 & 0\\
0 & 0 & 0\\
0 & 0 & a_2 \\
\end{pmatrix}.
\end{equation}

These cases could be called Semi-Parallel textures.
The explicit form of the 33 and 23 elements is  given by:

\begin{eqnarray}
 (\tilde{Y}^f_2)_{23} &=&  \frac{\sqrt{m_2 m_3}}{v}\chi_{23}, \\ \nonumber
 (\tilde{Y}^f_2)_{33} &=& \frac{m_3}{v}\chi_{33},
 \end{eqnarray}
where $\chi_{23}=\frac{\sqrt{ R}}{Q} \left(\frac{\sqrt{2} P}{\sin{\beta}}-\frac{a_1v}{m_3}\tan\beta\right)   \sqrt{\gamma_f}            $,  $\chi_{33}=\frac{R }{Q}\left(\frac{\sqrt{2} P}{\sin{\beta}}-\frac{a_1v}{m_3}\tan\beta \right)$ and $P=1-r_2 \gamma_f$.
In this case the elements depend on the  parameters $a_1$, $\beta$ and $\gamma_f$  which simplifies the
study of the phenomenology of the model. 

{{III) ``Complementary textures"}}  Here, we consider Yukawa matrices that 
have a different structure, but in such a way that they produce a hermitian mass matrix with
four zero textures. We shall consider patterns where at most one element from
each of the Yukawa matrices ${Y}^f_{1,2}$ contributes to one entry of the full  mass matrix. 

Thus the  cases that will be considered are defined as 
follows:\\

\begin{equation}
Y_1 = 
\begin{pmatrix}
0 & d & 0 \\
d^* & c & b \\
0 & b^* & 0
\end{pmatrix},
\hspace{5mm}
Y_2 = 
\begin{pmatrix}
0 & 0 & 0 \\
0 & 0 & 0 \\
0 & 0 & a
\end{pmatrix}.
\end{equation}

\begin{figure}[H]
\centering
    {\includegraphics[scale = 0.4]{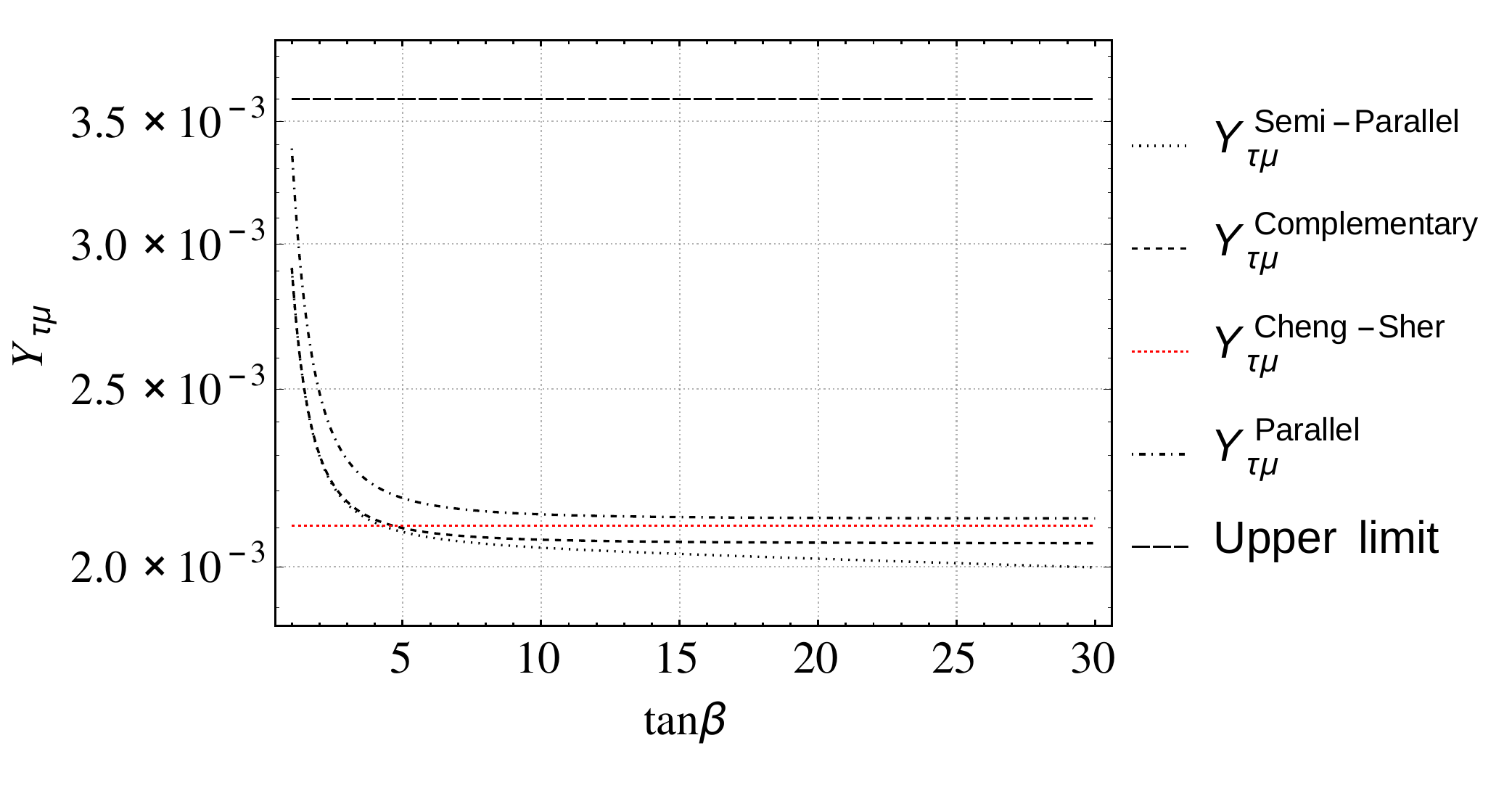}}       
   \caption{Plots of $Y_{\tau\mu}$($=\frac{\sqrt{m_{\tau}m_{\mu}}}{v}\chi_{\tau\mu}$) vs $\tan{\beta}$, with $\chi_{\tau\mu}$=1 for the Cheng-Sher ansatz and $\chi_{\tau\mu}\sim 1$ for the considered cases, i.e., Parallel, Semi-Parallel and Complementary. The highest line indicates the upper bound of the Yukawa coupling of $\tau$-$\mu$.  \label{ansatz}}
	\end{figure}

The explicit form of the 33 and 23 elements is  given by:

\begin{equation}
(\tilde{Y^f_2})_{23}=\frac{\sqrt{m_2m_3}}{v}\chi_{23},
\end{equation}
\begin{equation} 
 (\tilde{Y^f_2})_{33}=\frac{m_3}{v}\chi_{33},
 \end{equation}
where  $\chi_{23}=\frac{\sqrt{R}}{Q}\frac{\sqrt{2}P}{\sin\beta}\sqrt{\gamma_f}$ and $\chi_{33}=\frac{R}{Q}\frac{\sqrt{2}P}{\sin\beta}$. We can see that in the limit $a_1\to0$ the Semi-Parallel case is reduced to the Complementary case. In this case the elements depend only on the  parameters $\beta$ and $\gamma_f$  which is simplest case for the study of the phenomenology of the model because it has only two parameters.
 In Fig \ref{ansatz} a comparison is made between the Cheng-Sher ansatz and corrections arising 
 from the elements of $\chi_{ij}$ of order $\mathcal{O}(1)$ \cite{Cheng-Sher} which comes from the considered textures.

The example of ``Complementary textures" is called case 1, while the characteristics of other possible variations  (cases 2-6)
are included  the appendix \ref{Txt-cases2-6},  where some salient features are also mentioned.
In making these choices, we have assigned the 33 mass entry to $Y_2$. Our convention is motivated by 
the assumption that the largest Yukawa element is expected  to arise from some dominant mechanism (unknown)
 while the lightest masses and CKM mixing could be the result of some perturbation to the above mechanism.
The logic of our ordering is to start from the simplest structure for $Y_2$, i.e., with only $(Y_2)_{33} \ne 0$ (case 1), 
up to the more complicated case, where $Y_2$ itself is of the 3-texture type.

\section{\label{sec:level2} The Higgs Interaction in the THDM-Tx}

\subsection{The General Yukawa Lagrangian}

The Yukawa Lagrangian in the THDM-III is given by \cite{DiazCruz:2004pj, GomezBock:2005hc}

\begin{eqnarray}
\nonumber \mathcal{L}&=& {Y^u_1}\overline{Q}_{L}^0 {\hgt} u_{R}^0  + {{\yyuu}} \overline{Q}_{L}^0 {\hgtt} u_{R}^ 0 + {{\yd}} \overline{Q}_{L}^0 {\hggs} d_{R}^0 + {Y^d_2} \overline{Q}_{L}^0 {\hggss} d_{R}^0  \\
 &+& {{Y_1^l}} \overline{L}^0 {\hggs} l_R^0 + {Y_2^l} \overline{L}^0 {\hggss} l_R^0+h.c.,
\end{eqnarray}

such that
	
\begin{eqnarray}
\nonumber Q^0_L &=&  \left( \begin{array}{c}
u_L \\
d_L \\
        \end{array} \right),  \hspace{3mm}
L^0 =  \left( \begin{array}{c}
\nu_L \\
e_L \\
        \end{array} \right), \hspace{3mm}
        \Phi_1 =  \left( \begin{array}{c}
\phi_1^{+} \\
\phi_1^0 \\
        \end{array} \right),  \hspace{3mm}
        \Phi_2 =  \left( \begin{array}{c}
\phi_2^{+} \\
\phi_2^0 \\
        \end{array} \right), \\
        \tilde\Phi_j &=& i\sigma_2\Phi_j^{*} = \left( \begin{array}{c}
{\phi_j^{0}}^{*} \\
-\phi_j^{-} \\
        \end{array} \right).
\end{eqnarray} 

The physical fields are defined through a transformation that depends on the angle $\alpha$ which transforms the real part neutral physical Higgs bosons:

\begin{eqnarray}\label{matrizP}
\left( \begin{array}{c}
H^0            \\
h^0 \\
\end{array} \right) =
\left( \begin{array}{cc}
\cos\alpha & \sin\alpha             \\
-\sin\alpha & \cos\alpha \\
\end{array} \right)\cdot\left( \begin{array}{c}
Re\phi_1            \\
Re\phi_2 \\
\end{array} \right), 
\end{eqnarray}
and the angle $\beta$ which that transforms the imaginary part neutral and charged physical Higgs bosons given by

\begin{eqnarray}\label{matrizP}
\left( \begin{array}{c}
G^0            \\
A^0 \\
\end{array} \right) =
\left( \begin{array}{cc}
\cos\beta & \sin\beta             \\
-\sin\beta & \cos\beta \\
\end{array} \right)\cdot\left( \begin{array}{c}
Im\phi_1            \\
Im\phi_2 \\
\end{array} \right), 
\end{eqnarray}
\begin{eqnarray}\label{matrizP}
\left( \begin{array}{c}
G^{\pm}            \\
H^{\pm} \\
\end{array} \right) =
\left( \begin{array}{cc}
\cos\beta & \sin\beta             \\
-\sin\beta & \cos\beta \\
\end{array} \right)\cdot\left( \begin{array}{c}
\phi_1^{\pm}            \\
\phi_{\pm} \\
\end{array} \right), 
\end{eqnarray}
the angle $\beta$ is defined as
\begin{equation}
\tan\beta=\frac{v_2}{v_1}.
\end{equation}

As a matter of convenience we separate the Lagrangian into charged ($\dL_{ch}$) and neutral sector ($\mathcal{L}_n$). 
 Thus,

\begin{eqnarray}\label{lagrangiano}
\nonumber  \mathcal{L}_{n} &=& \frac{g}{2}\left(\frac{m_{d}}{m_{W}}\right)\bar{d}\left[\frac{\cos\alpha}{\cos\beta}\delta_{dd{'}}+\frac{\sqrt{2}\sin(\alpha-\beta)}{g\cos\beta}\left(\frac{m_{W}}{m_{d}}\right)\left(\tilde{Y}_{2}^{d}\right)_{dd{'}}\right]d{'}H^{0} \nonumber \\
 & + & \frac{g}{2}\left(\frac{m_{d}}{m_{W}}\right)\bar{d}\left[-\frac{\sin\alpha}{\cos\beta}\delta_{dd{'}}+\frac{\sqrt{2}\cos(\alpha-\beta)}{g\cos\beta}\left(\frac{m_{W}}{m_{d}}\right)\left(\tilde{Y}_{2}^{d}\right)_{dd{'}}\right]d{'}h^{0} \nonumber\\
 & + & \frac{ig}{2}\left(\frac{m_{d}}{m_{W}}\right)\bar{d}\left[-\tan\beta\delta_{dd{'}}+\frac{\sqrt{2}}{g\cos\beta}\left(\frac{m_{W}}{m_{d}}\right)\left(\tilde{Y}_{2}^{d}\right)_{dd{'}}\right]\gamma^{5}d{'}A^{0} \nonumber \\
 & + & \frac{g}{2}\left(\frac{m_{u}}{m_{W}}\right)\bar{u}\left[\frac{\sin\alpha}{\sin\beta}\delta_{uu{'}}+\frac{\sqrt{2}\sin(\alpha-\beta)}{g\sin\beta}\left(\frac{m_{W}}{m_{u}}\right)\left(\tilde{Y}_{2}^{u}\right)_{uu{'}}\right]u{'}H^{0} \nonumber\\
 & + & \frac{g}{2}\left(\frac{m_{u}}{m_{W}}\right)\bar{u}\left[-\frac{\cos\alpha}{\sin\beta}\delta_{uu{'}}+\frac{\sqrt{2}\cos(\alpha-\beta)}{g\sin\beta}\left(\frac{m_{W}}{m_{u}}\right)\left(\tilde{Y}_{2}^{u}\right)_{uu{'}}\right]u{'}h^{0} \nonumber \\
  &+ & \frac{ig}{2}\left(\frac{m_{u}}{m_{W}}\right)\bar{u}\left[-\cot\beta\delta_{uu{'}}+\frac{\sqrt{2}}{g\sin\beta}\left(\frac{m_{W}}{m_{u}}\right)\left(\tilde{Y}_{2}^{u}\right)_{uu{'}}\right]\gamma^{5}u{'}A^{0}.
\end{eqnarray}
The lepton part is similar to the type-down quarks part with the exchange $d\to l$ and $m_d\to m_l$.
	
We will use a notation in which $\eta_{f\bar{f}}^H$ corresponds to the couplings $f\bar{f}H$ in the Lagrangians, for example:
\begin{equation}
\eta_{l l^{'}}^{H^0}=\frac{g}{2}\left( \frac{m_l}{m_W} \right)  \left[\frac{\cos\alpha}{\cos\beta} \delta_{ll{'}} + \frac{\sqrt{2}\sin(\alpha-\beta)}{g\cos\beta} \left( \frac{m_W}{m_l} \right) \left( \tilde Y_2^l \right)_{ll{'}}\right]\label{NeuLag}.
\end{equation}

While the charged Yukawa Lagrangian is given by:
\begin{eqnarray}\label{lagrangianoCH}
\nonumber \mathcal{L}_{ch} & = & \left[\bar{d}_{i}\left(\tilde{Y}_{1}^{u}\sin\beta+\tilde{Y}_{2}^{u}\cos\beta\right)u_{j}H^{-} + \bar{u}_{i}\left(\tilde{Y}_{2}^{d}\cos\beta-\tilde{Y}_{1}^{d}\sin\beta\right)d_{j}H^{+}\right]P_{R}\\
 & + & \left[\bar{d}_{i}\left(\tilde{Y}_{2}^{u}\cos\beta-\tilde{Y}_{1}^{u}\sin\beta\right)u_{j}H^{-} + \bar{u}_{i}\left(\tilde{Y}_{1}^{d}\sin\beta-\tilde{Y}_{2}^{d}\cos\beta\right)d_{j}H^{+}\right]P_{L}. 
\end{eqnarray}
from which we extract the Feynman rules for the charged Higgs mediated processes.

\section{Constraints from Low Energy}\label{sec:cons-low-ener}

In order to find the allowed regions of parameter space, we will consider $\Delta a_{\mu}^{THDM-Tx}$, relevant low energy processes and collider constraints.
Unlike the SM we have an additional doublet, which gives rise to new Feynman diagrams mediated by $h^0,\,H^0,\,A^0,\,H^{\pm}$. 
Using current measurements we constrain the parameter space of our model which depends on $\tan{\beta}$ which 
mediates the coupling of the Higgs bosons to fermions. We will determine the parameter space analyzing the $t_{\beta}-m_{H(H^{\pm})}$ plane, 
while the other parameters remain fixed. In regard to the angles $\alpha$ and $\beta$ we study the scenario $(\alpha-\beta) = \frac{\pi}{2}$, 
which reproduces the case where the coupling of $h^0$ to fermions is SM-like, also because it is the most favorable scenario according 
to \cite{a-b}. For process mediated via neutral Higgs bosons, we consider anomalous magnetic dipole moment $(\Delta a_{\mu}^{THDM-Tx})$, 
$B_s^0 \to \mu^+\mu^-$, $\mu\rightarrow e \gamma$, $\tau \rightarrow \mu \gamma$, $\tau \rightarrow e \gamma$, $K-\bar{K}$ $(B-\bar{B})$ mixing 
and $\tau\rightarrow \mu^-\mu^+\mu^-$. But for charged Higgs mediated processes we employ $\Delta a_{\mu}^{THDM-Tx}$, $B\to D(D^*)\tau\nu_{\tau}$ and $\mu\to e\gamma$.

We use the effective Hamiltonian formalism in order to analyze the low energy processes, which is based on the expansion 
\begin{equation}
\mathcal{H}_{eff} = \frac{G_F}{\sqrt{2}}\sum_i V_{CKM}^i C_i(\mu)Q_i,
\end{equation}
where $Q_i$ are the effective operators that govern the decay and $C_i(\mu)$ the Wilson Coefficients; $\mu$ separates the physics energy scale 
contributions into high and low energy scales. The Wilson Coefficients are dependent on the couplings of our model, and so they depend 
on $\alpha$, $\beta$, $\alpha_i$, $m_{H(H^{\pm})}$ and $\gamma_f$, from which we are able to constrain our parameter space. 
\subsection{ $ B_s^0 \to \mu^+\mu^-$}

\begin{figure}[H]
\centering
\includegraphics[scale = 0.5]{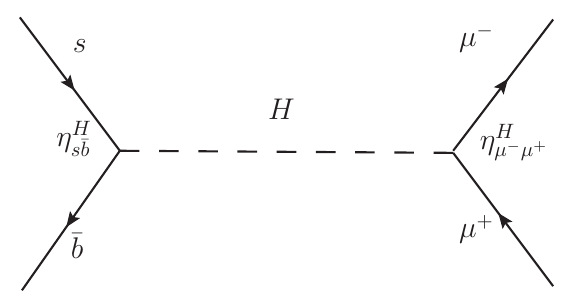}
\caption{ Feynman diagram for the $B_s^0\to\mu^+\mu^-$ process at the tree level, where $H=h^0,\,H^0,\,A^0$. The labels $\eta_{f\bar{f}}^H$ indicate the contributions coming from the THDM-Tx. We omit these labels in all further Feynman diagrams.
\label{fd:Btomumu}}
\end{figure}

$B_s^0$ meson decays into charged muons $\mu^+\mu^-$ are very interesting due to their sensitivity to BSM theories.
One of which is the THDM; in this scenario the process $B_s^0 \to \mu^+\mu^-$, is mediated by a neutral Higgs boson, 
whose Feynman graph at the quark level is shown in Fig. \ref{fd:Btomumu}.

This decay is particularly important because impose stringent constraints on several theories beyond the SM. The first observation of the rare $B_s^0\to\mu^{-}\mu^{+}$ decay at the LHC by the CMS and LHCb experiments \cite{Khachatryan} reported a result of its branching ratio, whose value is 

\begin{equation}
 BR(B_s^0\rightarrow\mu^+\mu^-)=(2{.}8_{-0{.}6}^{+0{.}7})\times10^{-9},
\end{equation}
while the SM value reported by \cite{BmumuSM} is
\begin{equation}
 BR(B_s^0\rightarrow\mu^+\mu^-)_{SM}=(3{.}23\pm0{.}27)\times10^{-9}.
\end{equation}
\nolinebreak

The branching ratio for this decay is given by \cite{dedes},

	\begin{eqnarray}
 BR\left(B_{s}\rightarrow\mu^+\mu^-\right) &=&
\frac{G_{F}^{2}\alpha_{em}^{2}}{16\pi^{3}}M_{B}\tau_{B}\left|V_{ts}V_{tb}^{*}\right|^{2}\sqrt{1-\frac{4m_{\mu}^{2}}{M_{B}^{2}}}\nonumber \\ &\times & \left[\left|F_{RH}\right|^{2}\left(1-\frac{4m_{\mu}^{2}}{M_{B}^{2}}\right)+\left|F_{RA^0}\right|^{2}\right],
 	\end{eqnarray}

\noindent where

	\begin{equation} 
F_{RH,RA^0}=-\frac{i}{2}M_{B}^{2}\, f_{B_{b}}\frac{m_{b}}{(m_{b}+m_{s})m_{\mu}}C_{RH,RA^0},
 	\end{equation}

\noindent are the form factors, $ G_{F}$ is the Fermi constant, $\tau_{B}$ is the lifetime of the $B$ meson, $f_{B_{b}}$ is the meson decay constant and $C_{RH}$, $C_{RA^0}$ 
are the Wilson coefficients  that  appear in the effective Hamiltonian,

\begin{equation}
H_{eff}=-2\sqrt{2}G_{F} V_{tb}V_{tb}^{*}\sum_i C_i Q_i,
\end{equation}

\noindent where $Q_i$ are the effective operators with $i=RH, RA^0$:

	\begin{eqnarray} 
Q_{RH}&=& \frac{e^{2}}{16\pi^{2}}(\bar{s\,}P_{R}b)({\mu^+}\, \mu^-), \\ 
Q_{RA^0}&=&\frac{e^{2}}{16\pi^{2}} (\bar{s\,}P_{R}b)({\mu^+}\gamma^{5}\mu^-), 
	\end{eqnarray}

\noindent and the Wilson coefficients $C_i$ are,

\begin{eqnarray}
 C_{RH} &=&  \frac{2\pi m_{\mu}^2}{V_{ts}^{*}V_{tb}\alpha_{em}}\Bigg[\frac{1}{4m_{H^{0}}^{2}}\sum_{H=H^0,\,h^0}\eta_{\bar{b}s}^H \eta_{\mu^{-}\mu^{+}}^H
\Bigg], \label{1}
	\end{eqnarray}	
	
	\begin{eqnarray}
C_{RA^0}=\frac{2\pi m_{\mu}^2}{V_{ts}^{*}V_{tb}\alpha_{em}} \Bigg[\frac{1}{4m_{A^{0}}^{2}} \eta_{\bar{b}s}^{A^0} \eta_{\mu^{-}\mu^{+}}^{A^0}
 \Bigg].\label{2}
	\end{eqnarray}
\nolinebreak Here $m_b,\,V^*_{tb(ts)}, M_{H^0}, M_{h^0}, M_{A^0}$ are the
bottom quark mass, CKM elements and Higgs boson masses,
respectively. Operators with Left-handed b-quarks are parametrically suppressed compared to those involving Right-handed by a factor $m_s$/$m_b$.
Therefore, they are sub-leading corrections for our work. However there are some cases where these corrections are important, for instance
when you have a complete general flavor violating MSSM. See for example section II of Ref.\cite{dedes} for the complete effective Hamiltonian and the Wilson coefficients.We observe from \ref{1}-\ref{2} the dependence on the model parameters.

\subsection{$a_{\mu}^{THDM-Tx}$}
 There currently exists a discrepancy between the experimental measurement and the theoretical prediction of $a_\mu$ in the SM \cite{MomMagMuonEXP}
\begin{eqnarray}
\Delta a_\mu = a_\mu^{Exp} - a_\mu^{SM} = 288(63)(49)\times 10^{-11}, 
\end{eqnarray}
which is greater than 3$\sigma$, this discrepancy might originate from physics beyond the SM. That is why it is important to implement it to constrain new models, 
such as in the case of the THDM-Tx; in this model new contributions to $a_{\mu}$ come from one the loop level couplings $h^0\mu l$, $H^0\mu l$, $A^0\mu l$, 
where $l=e,\,\mu,\,\tau$.

\begin{figure}[H]
\centering
    \subfigure[ ]{\includegraphics[scale = 0.37]{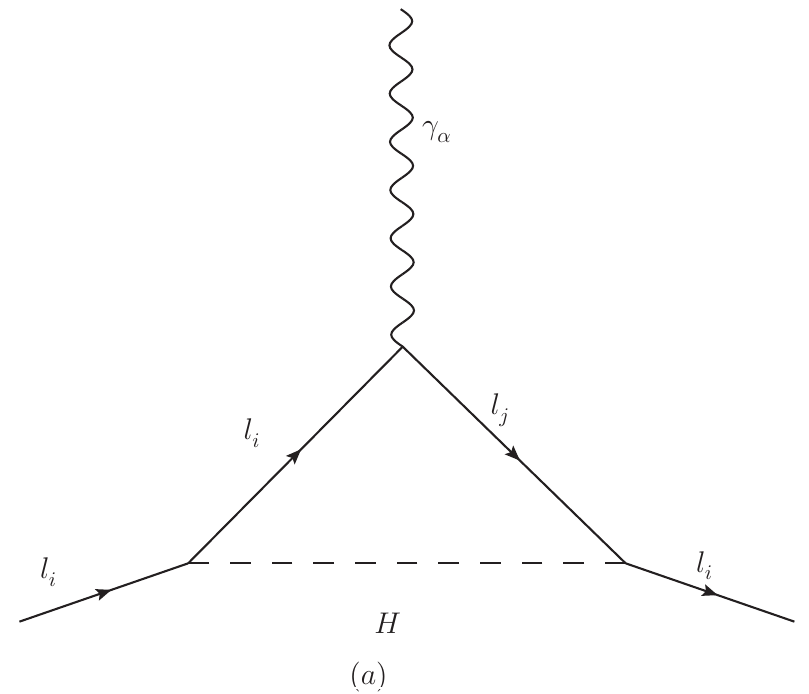}}
    \subfigure[ ]{\includegraphics[scale = 0.37]{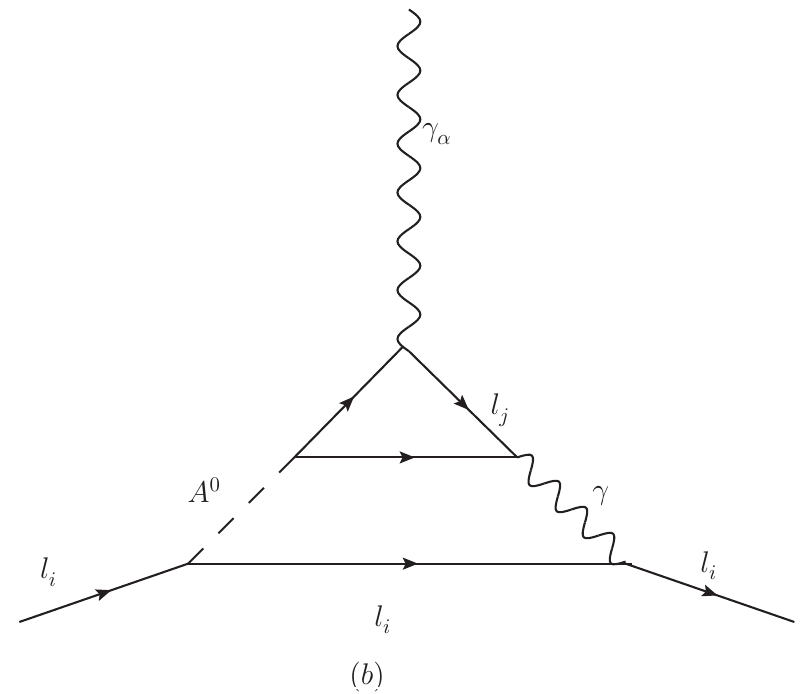}}    
    \subfigure[ ]{\includegraphics[scale = 0.37]{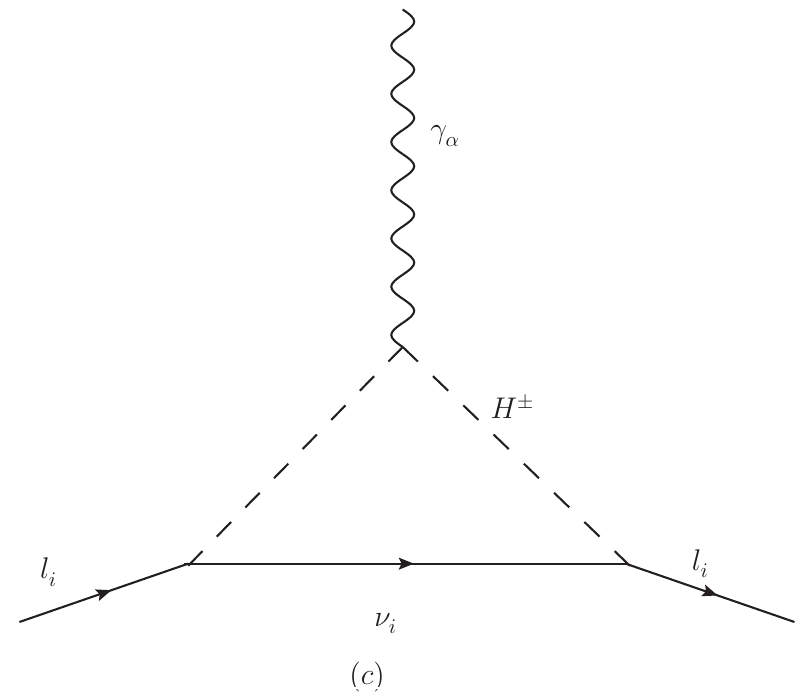}}     
   \caption{Feynman diagrams of the muon anomalous magnetic dipole moment.  \label{FD}}
	\end{figure}

The Feynman diagrams for these one loop level processes are shown in FIG. \ref{FD}
 
\noindent Where $l_i=\mu$ and $l_j$ is a different flavored lepton, $H$ refers to the different neutral Higgs bosons $h^0, H^0, A^0$. 
The  muon anomalous magnetic dipole moment is CP-conserving, thus $\alpha_{1,\,2}=0$.

We use the unitary gauge and the method of Feynman parametrization. The expressions for the $a_{\mu}^{THDM-Tx}$, at one-loop level, are given by:

\begin{equation}
 a_{\mu}^{THDM-Tx}=\underset{\underset{l'=\mu}{l=e,\,\mu,\, \tau}}{\sum}\frac{\left|\eta_{_{ll'}}^{H}\right|^2 m_{\mu}}{\sqrt{8}\pi^{2}}\intop_{0}^{1}dx\intop_{0}^{1-x}dy\,F_{k}(x,\, y),
\end{equation}
\nolinebreak
where $\eta_{_{ll'}}^{H}$ is the coupling $l\bar{l}H$ that comes from the neutral Lagrangian, Eq. \ref{NeuLag}. \\

The $F_{k}(x,\, y)$ function of diagram (a) is given by:

\begin{equation}\label{eq1}
F_{a}(x,\,y)=(x+y)(m_{l_{j}}-m_{\mu}(x+y-1))/M_{a}^{2},
\end{equation}
\nolinebreak
where $M_{a}^{2}=-m_{H}^{2}(x+y-1)+(x+y)(m_{l_{j}}^{2}+m_{\mu}^{2}(x+y-1)$.

 For diagram (c): $\eta_{_{ll'}}^{H}=1$ and 
\begin{equation}\label{eq2}
F_{c}(x)=2m_{\mu}x/M_{c}^{2},
\end{equation}
\nolinebreak
such that $M_{c}^{2}=(m_{\mu}^{2}x-m_{H^{\pm}}^{2})$. \\
In the case of the Barr-Zee diagram, we only considerer the dominant contribution which is given by \cite{Barr-Zee}:
\begin{equation}
a_{\mu}^{Barr\,Zee}=\frac{\alpha^2}{8\pi^2s_W^2}\frac{m_{\mu}^2 \eta_{\mu^{-}\mu^{+}}^{A^0}}{m_W^2} \sum_{f=t,\,\tau,\,b} N_c^fQ_f^2r_ff(r_f)\eta_{f\bar{f}}^{A^0},
\end{equation}
where $r_f=(m_f/M_{A^0})$, $m_f$ is the fermion mass, $N_c^f=1(3)$ is the color number: in the case of leptons (quarks), 
$Q_f$ is the electric charge of fermions, $\eta_{f\bar{f}}^{A^0}$ is given by the Lagrangian in eq.\ref{lagrangiano} and the function $f(r_f)$ is given by
\begin{equation}
f(x)=\int_0^1 \frac{log(\frac{x}{y(1-y)})}{x-y(1-x)}dy.
\end{equation}


\subsection{\textnormal{Meson Mixing: $K\bar{K}$ and $B\bar{B}$ Mixing}}

\begin{figure}[H]
\centering
\includegraphics[scale = 0.5]{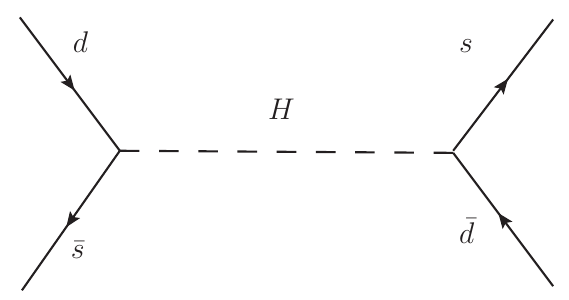}
\caption{ Feynman diagram for the $K-\bar{K}$ mixing.
\label{fd:sd-phi-ds}}
\end{figure}

One of the earliest successful tests of SM phenomenology was the measurement of the difference between the mass eigenvalues 
$\Delta M_K = M_L - M_S$; related to $M_{1,2}^K$ which occur through box diagrams involving the W boson. But in the THDM the 
amplitude of $K\bar{K}$ mixing receives additional contributions from the neutral scalars, whose Feynman diagram is shown in Fig. \ref{fd:sd-phi-ds}.

The effective hamiltonian for this process at the quark level is written as follows

\be
 H_{eff}^{\Delta S =2} = \frac{G_F^2 M_W^2}{16\pi^2}\sum_i C_i Q_i,
\ee

with $(i=L, R, LR)$ 

\be
 Q_L^{\overline{s}d,s\overline{d}} &=&  \big(\bar{s} P_L d\big) \big(\bar{s} P_L d\big),\\
 Q_R^{\overline{s}d,s\overline{d}} &=&  \big(\bar{s} P_R d\big) \big(\bar{s} P_R d\big),\\
 Q_{LR}^{\overline{s}d,s\overline{d}} &=&  \big(\bar{s} P_L d\big) \big(\bar{s} P_R d\big),
\ee

and the Wilson coefficients are written here for a general multi-Higgs doublet model,

\be
C_L^{\overline{s}d,s\overline{d}} = - \frac{16 \pi^2}{G_F m_W^2}m_sm_d |\eta_{sd}^H|^2 \sum_{a=1,\, H}^2 \frac{\big(U_{1a}^*\big)^2 }{m_{H}^2}g,\\
C_R^{\overline{s}d,s\overline{d}} = - \frac{16 \pi^2}{G_F m_W^2}m_sm_d |\eta_{sd}^H|^2 \sum_{a=1,\, H}^2 \frac{\big(U_{1a}\big)^2 }{m_{H}^2},\\
C_{LR}^{\overline{s}d,s\overline{d}} = - \frac{16 \pi^2}{G_F m_W^2}m_sm_d |\eta_{sd}^H|^2 \sum_{a=1,\, H}^2 \frac{U_{1a}^* U_{1a} }{m_{H}^2},
\ee

the rotation matrix for neutral Higgs bosons takes the form

\be
U= 
 	\begin{pmatrix}
\cos\alpha&-\sin\alpha\\
\sin\alpha&\cos\alpha\\
  	\end{pmatrix}.
\ee

Flavor violation arises from the non-diagonal terms of the K (neutral kaon) mass matrix, in particular the component 
$M_{12}^K,$  whose experimental value has been measured to be

\be
M_{12}^K = \frac{\Delta M_K}{M_K} = 7.2948 \times 10^{-15},
\ee

\noindent which is a high precision measurement that is used to constrain BSM parameters. Given that $\Delta M_K$ is obtained through:

	\begin{eqnarray}
 \Delta M_K &=& 2 Re\langle \bar{K}^0|H_{eff}^{\Delta S = 2}|K^0\rangle \\ \nn
&=& \frac{G_F^2 M_W^2}{12 \pi^2} M_K F_K^2 \eta_2 {B}_K  
\times\left[ \bar{P}_{2,LR} C_{LR}^{\overline{s}d,s\overline{d}}  
+ \bar{P}_{1,L} \left(C_L^{\overline{s}d,s\overline{d}} + C_R^{\overline{s}d,s\overline{d}}\right) \right] ,
	\end{eqnarray}

\noindent where $F_K = 160 $ MeV, $M_K = 497.6$ MeV, $ \eta_2 = 0.57, \;\; {B}_K = 0.85\pm 0.15,\;\; \bar{P}_{2,LR} = 30.6$ 
and $\bar{P}_{1,L} = -9.3,$ therefore,

	\begin{eqnarray}
 &M&_{12}^K = \frac{4}{3} F^2_K \eta_2 \bar{B}_K \big(m_d m_s\big) \frac{1}{v_1^2} \times \sum_{a = 1}^3 \bigg[
 \bar{P}_{2,LR} \frac{U_{2a} U_{1a}}{m_{H}^2} 
                 + \bar{P}_{1,L}\bigg(\frac{U_{2a}^2}{m_{H}^2} + \frac{U_{1a}^2}{m_{H}^2}\bigg)
\bigg]. 
	\end{eqnarray} \\
	
The way in which we determine $B-\bar{B}$ mixing is analogous to $K\bar{K}$ mixing, except that we now use $\Delta M_B$. For this process the 
experimental values used are $\Delta m_{B_s} = 3.337 \times 10^{-13}$ GeV, $\eta_B = 0.55, \;\; \bar{P}_{2,LR} = 0.88$ and $\bar{P}_{1,L} = -0.52$.

 \subsection{$B\rightarrow D\tau\nu$ and 
$B\rightarrow D^\star\tau\nu$}

\begin{figure}[H]
\centering
\includegraphics[scale = 0.5]{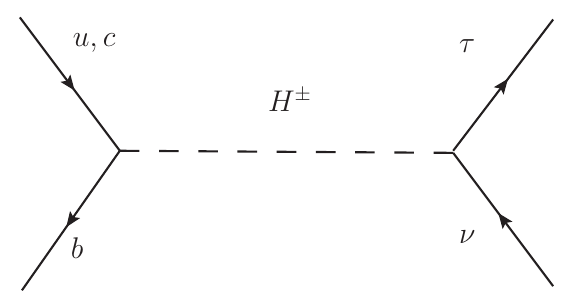}
\caption{ Feynman diagram for the $B\rightarrow D(D^*)\tau\nu$ process.
\label{fd-u_c_b-nu_tau}}
\end{figure}

The experiments Belle and BaBar have now measured the ratios 
R($D$) and R($D^\star$) \cite{Aubert:2009at, Matyja:2007kt}, of $B\rightarrow D\tau\nu$ and $B\rightarrow D^\star\tau\nu$; which can be used to 
constrain charged Higgs $H^\pm$ parameters appearing in models such as in the case of the THDM-Tx.
The Feynman diagram corresponding to this process is shown in Fig. \ref{fd-u_c_b-nu_tau},

the results given by BaBar are:

\begin{eqnarray}
R(D) &=& 0.44 \pm 0.058 \pm 0.042, \\ \nonumber
R(D^*) &=& 0.332 \pm 0.024 \pm 0.018, \nonumber
\end{eqnarray}

which can be expressed as follows,
\begin{equation}
{R}(D) = {R}_{ SM}(D)\Big{(} 1+ 1.5 \;{\mbox{Re}}\left[\frac{C_{R}^{cb\,, \tau\nu }+C_L^{cb\,, \tau\nu}}{C_{SM}^{cb \,, \tau\nu}}\right] \\ \nn
+ 1.0  \left| \frac{C_{R}^{cb\,, \tau\nu}+C_L^{cb\, ,\tau\nu}}{C_{SM}^{cb \,, \tau\nu}}\right|^2 \Big{)} \,, \label{RD}  
\end{equation}
\begin{equation}
{R}(D^*) = {R}_{ SM}(D^*) \Big(1 + 0.12\; {\mbox{Re}} \left[\frac{C_{R}^{cb\,, \tau\nu}-C_L^{cb\,, \tau\nu}}{C_{SM}^{cb \,, \tau\nu }}\right] \
+0.05 \left|\frac{C_{R}^{cb\, ,\tau\nu}-C_L^{cb\, ,\tau\nu}}{C_{SM}^{cb \,, \tau\nu}}\right|^2  \Big{)}\,,
\end{equation}

\noindent where the Wilson coefficients are given in Ref. \cite{Crivellin:2012ye}.

The combination of both processes give us a deviation of 3.4 $\sigma$ with respect to SM predictions; it remains to be seen if this is in fact a signal of new physics.

This signal is one of the most stringent processes. Nonetheless, we are able to satisfy both 
$R(D)$ and $R(D^\star)$ within of THDM-Tx. The allowed and excluded regions for $\tan\beta$-$M_{H^{\pm}}$ plane are presented in \ref{regionespermitidas}

 \subsection{\textnormal{Lepton decays $ l_i\to l_j\gamma$ }}
Another way in which one may constrain the parameter space of a model is through the consideration of radiative flavor violating decays, 
such as in the case of leptonic decays of which $\mu\to e\gamma$ is particularly useful. The MEG collaboration \cite{Adam:2013mnn} has 
given an upper bound for the decay $\mu\to e\gamma$: $5.7\times 10^{-13}$, they also give bounds of the leptonic decays of the $\tau$ 
but they are much weaker: BR$(\tau\to e\gamma)$=$3.3\times 10^{-8}$ and BR$(\tau\to\mu\gamma)$=$4.43\times 10^{-8}$ \cite{Aubert:2009ag}. 
That is why in the present work we only used $\mu\to e\gamma$ to constrain our parameter space. 

\begin{figure}[H]
\centering
\includegraphics[scale = 0.5]{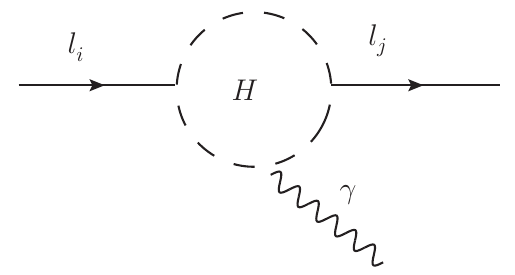}
\caption{Feynman diagram for the $l_i \to l_j \gamma$ process with $H=H^0, h^0, A^0, H^{\pm}$. The circle denotes one-loop contributions.
\label{fd:li-lj}}
\end{figure}

The Feynman diagram for this process is shown in Fig. \ref{fd:li-lj},

The branching ratio for the general decay of a lepton ($l_i$) to a lepton of a different family ($l_j$) is given by

\begin{eqnarray}
BR(l_i\to l_j\gamma) = \frac{m_{l_i}^5}{4\pi\Gamma_{l_i}}\left(\big|C_R^{l_jl_i}\big|^2 + 
\big|C_L^{l_jl_i}\big|^2\right),
\end{eqnarray}  
where $\Gamma_{l_i}$ is the total decay width of the particle $l_i$ and the Wilson Coefficients $C_{R,L}^{l_j l_i}$ are given by

\begin{eqnarray}
\nonumber C_{R}^{l_j l_i} &=&\sum_H\frac{-e}{192\pi^2M_H^2} \\
& &\Bigg[
\eta_{l_il_j}^{LRH\star} \eta_{l_il_j}^{LRH} + 
\eta_{l_jl_i}^{LRH\star} \eta_{l_jl_i}^{LRH} 
- \frac{m_{l_i}}{m_{l_j}}\eta_{l_il_j}^{LRH}\eta_{l_jl_i}^{LRH}
\left(9 + 6\ln\left(\frac{m_{l_j}^2}{m_{H}^2}\right) \right) \Bigg],
\end{eqnarray}
such that $H=h^0, H^0, A^0$. $C_L^{l_jl_i}$ is obtained by simply interchanging R by L. For the charged Higgs boson contributions the Wilson Coefficients are given by:
\begin{eqnarray}
C_{L}^{l_{j}l_{i}} & = & \frac{e}{384\pi^{2}M_{H^{\pm}}^2}{\sum_{k=1}^3}\xi_{\nu_{k}l_{i}}^{L}\xi_{\nu_{k}l_{j}}^{L},\\
C_{R}^{l_{j}l_{i}} & = & \frac{m_{l_{j}}}{m_{l_{i}}}\frac{e}{384\pi^{2}M_{H^{\pm}}^2}{\sum_{k=1}^3}\xi_{\nu_{k}l_{i}}^{R}\xi_{\nu_{k}l_{j}}^{R},
\end{eqnarray}
where

\[
\xi_{\nu_{k}l_{i}}^{L,R}=-{\sum_{m=1}^3}\sin\beta\,V_{km}^{PMNS}(\epsilon_{mi}\tan\beta),
\]
with
\[
|\epsilon_{mi}|\leq\left(\begin{array}{ccc}
2.9\times10^{-6} & 6.1\times10^{-4} & 1.0\times10^{-2}\\
6.1\times10^{-4} & 6.1\times10^{-4} & 1.0\times10^{-2}\\
1.0\times10^{-2} & 1.0\times10^{-2} & 1.0\times10^{-2}
\end{array}\right).
\]

\subsection{$\tau^-\rightarrow \mu^-\mu^+\mu^-$} 

\begin{figure}[H]
\centering
\includegraphics[scale = 0.5]{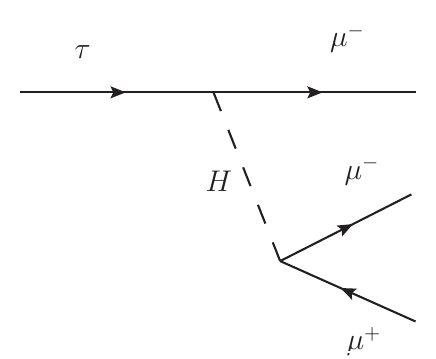}
\caption{ Feynman diagram for the $\tau^-\rightarrow \mu^-\mu^+\mu^-$ process.
\label{fd:liljlklk}}
\end{figure}

The Feynman diagram is shown in the FIG.[\ref{fd:liljlklk}],

The branching ratio including contributions from the three Higgs bosons is given by: 

\begin{eqnarray}
\nonumber BR(l_{i} & \rightarrow & l_{j}l_{k}{l}_{k})=\frac{5\delta_{l_{l}l_{k}}+2}{3}\frac{\tau_{i}}{2^{11}\pi^{3}}\frac{m_{l_{j}}m_{l_{k}}^{2}m_{l_{i}}^{6}}{v^{4}}\left\{ \frac{\cos^{2}(\alpha-\beta)sin^{2}\alpha}{m_{h^{0}}^{4}}\right.\\ \nonumber
 & + & \frac{\sin^{2}(\alpha-\beta)\cos^{2}\alpha}{m_{H^{0}}^{4}}-2\frac{\cos(\alpha-\beta)\sin(\alpha-\beta)\,\cos\alpha\, \sin\alpha}{m_{h^{0}}^{2}m_{H^{0}}^{2}}\\ 
 & + & \left.\frac{\sin^{2}\beta}{m_{A^{0}}^{4}}\right\} \frac{\left|\eta_{{ij}}\right|^{2}}{2\cos^{4}\beta}.
\end{eqnarray}
Here $\tau_{i}$ is the time life of the $l_i$ particle. The current bound is \cite{PDG:2016}:
 \begin{eqnarray}
BR(\tau^{-}\rightarrow\mu^{-}\mu^{+}\mu^{-}) & < & 2{.}1\times10^{-8}. \nonumber
\end{eqnarray}

\subsection{\textnormal{Allowed regions}\label{regionespermitidas}}

\subsubsection{Neutral Higgs Mediated Process}

\begin{figure}[h!]
 \centering
 {\includegraphics[scale = 0.25]{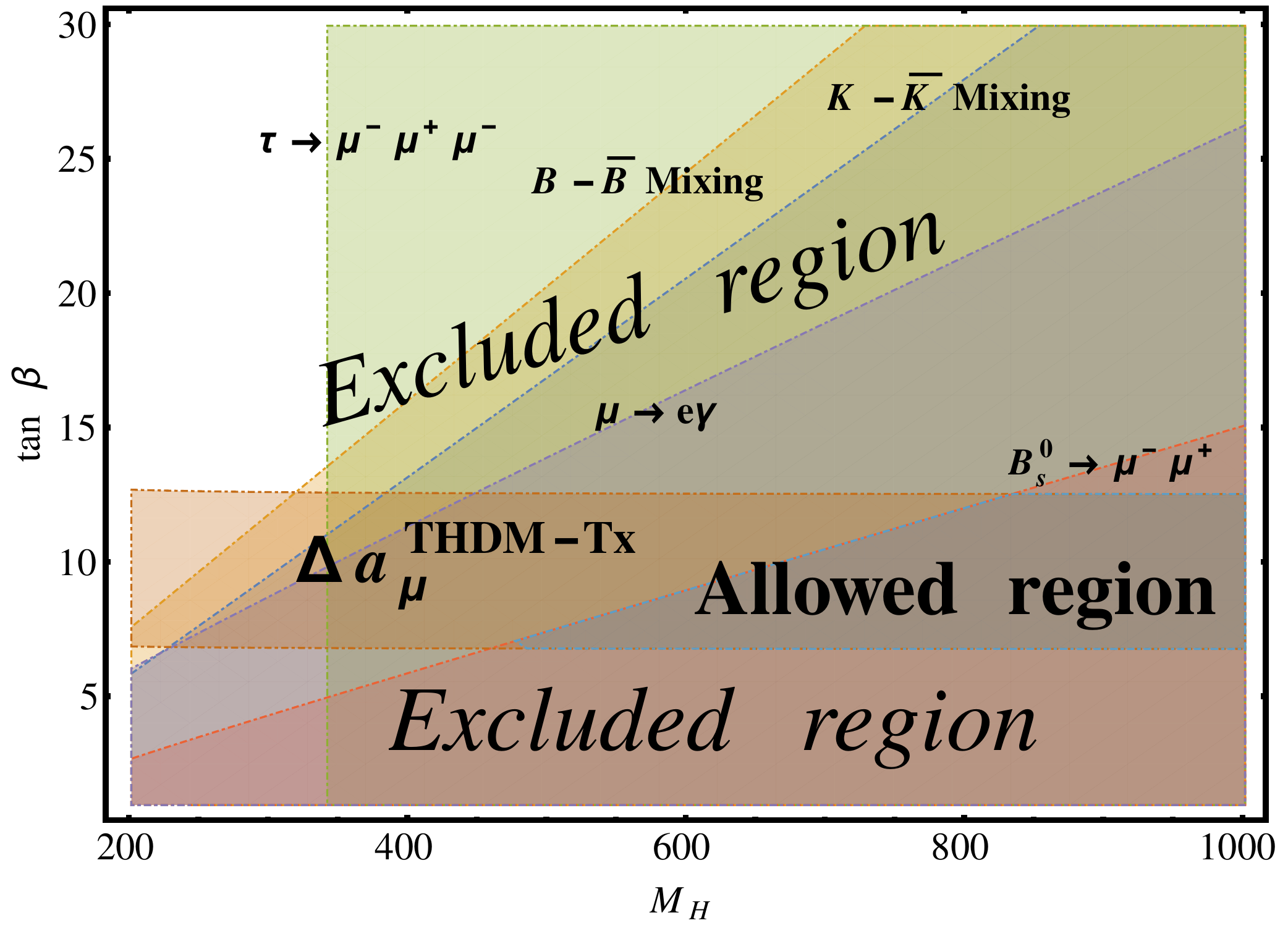}}
 \caption{Allowed and excluded regions of $t_{\beta}$ and $M_H$(GeV)  for the complementary case. \label{neutrosComple}}
	\end{figure}

\begin{figure}[h!]
 \centering
{\includegraphics[scale = 0.25]{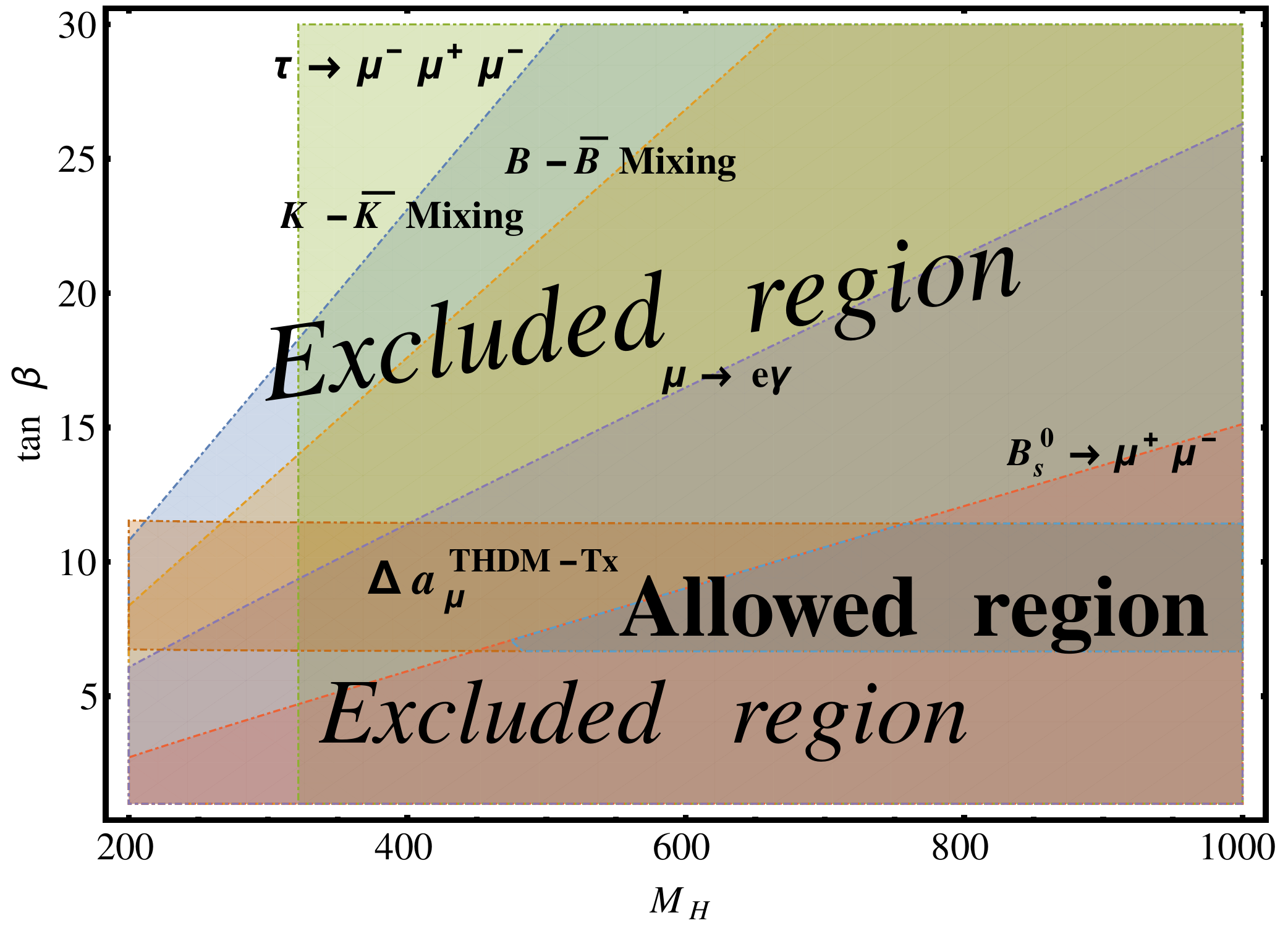}}
              \caption{We now perform a similar analysis  as in Fig. \ref{neutrosComple} but for the Semi-Parallel case. \label{neutrosSemi}}
	\end{figure}

\begin{figure}[h!]
 \centering
{\includegraphics[scale = 0.25]{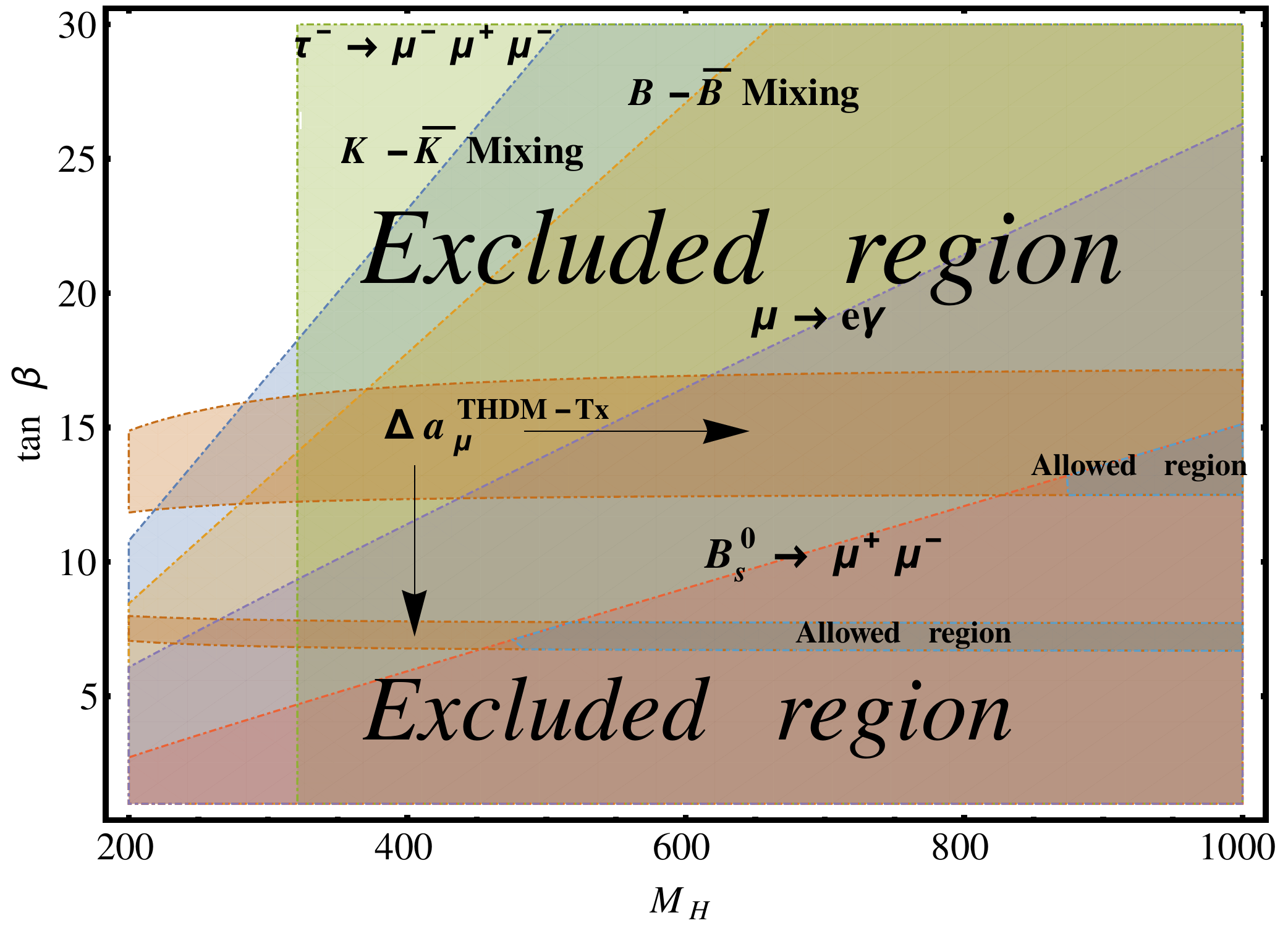}}
              \caption{Allowed and excluded regions for the Parallel case.  \label{Paralela}}
	\end{figure}

We give bounds on the $\tan\beta$-$M_H$ plane. We find that not all values of $M_H$ and $t_\beta$ are allowed. In Fig.\ref{neutrosComple} 
we show the allowed and excluded regions for processes that involve the neutral Higgs bosons for the 
complementary case, i.e. $B_s^0\to\mu^-\mu^+$, $K-\bar{K}(B-\bar{B})$ mixing, $l_i\to l_j\gamma$, $\tau^-\rightarrow \mu^-\mu^+\mu^-$ and $a_{\mu}^{THDM-Tx}$, 
where the region where they all intersect corresponds to the allowed region for all the process that we considered. 
We work in the scenario where $(\alpha-\beta)=\pi/2$ which is in accordance with the work of Ref. \cite{a-b}. 
In Fig.\ref{neutrosSemi} we consider the same low energy processes but for the Semi-Parallel case. 
Finally in Fig. \ref{Paralela} we show the excluded and allowed region for the parallel case.
	The labels for each process are placed at the upper limits for each process, which extends all the way down to ($M_H=200$GeV-$\tan{\beta}=1$)-($M_H=1000$GeV-$\tan{\beta}=1$). For example, for the process $B_s^0\to\mu^{-}\mu^{+}$ the allowed region starts at ($M_H=200$GeV, $\sim \tan{\beta}=3$) and continues until ($M_H=1000$GeV, $\sim \tan{\beta}=15$) and extends up to ($M_H=200$GeV, $\tan{\beta}=1$)-($M_H=1000$GeV, $\tan{\beta}=1$). The allowed regions that are different are:  $a_{\mu}^{THDM-Tx}$ which is a band which goes through the plot and $\tau^{-}\to\mu^{-}\mu^{-}\mu^{+}$ which looks like a rectangle. Here $M_H=M_{H^0}$ is the heavy Higgs boson mass.  We used the values $M_{h^0}=125$ GeV,   $M_{A^0}=300$ GeV, $M_{H^\pm}=500$ GeV, $\gamma_u=0.13$, $\gamma_d=0.1$ and $\gamma_l=1$. From the plot we observe that the most restrictive process in the $\tan{\beta}$-$M_H$ plane are $B_s^0\to\mu^{-}\mu^{+}$ and $a_{\mu}^{THDM-Tx}$. The intersected area for all the processes corresponds to the allowed region, which is found to be between $490\lesssim M_H\leq1000$ GeV for $t_{\beta}\sim7$ and $850\lesssim M_H\leq 1000$ for $\tan{\beta}\sim13$. 
	 Although both complementary and Semi-Parallel cases are similar, we observe that the allowed regions differ slightly, this is due to in the Semi-Parallel case there is an additional parameter ($a_1$). The most restrictive processes are ($B_s^0\to\mu^{-}\mu^{+}$ and $a_{\mu}^{THDM-Tx}$) just like in the previous plot. Nonetheless, the overall allowed region is greater than the complementary case, in particular for $t_\beta\sim13$ the Heavy Higgs boson mass is between  $750\lesssim M_H \leq 1000$ GeV unlike the complementary case is between  $850\lesssim M_H\leq 1000$ GeV.
	We find that there are two zones for the allowed parameter space: the first region is found at $7\lesssim \tan{\beta}\lesssim8$ for $480\lesssim M_H\leq 1000$ GeV while the second region is located at $12\lesssim \tan{\beta}\lesssim 15$ for $870\lesssim M_H\leq 1000$ GeV.
	
\subsubsection{Charged Higgs Mediated Process}
\begin{figure}
\centering
\subfigure[ ]{\includegraphics[scale = 0.16]{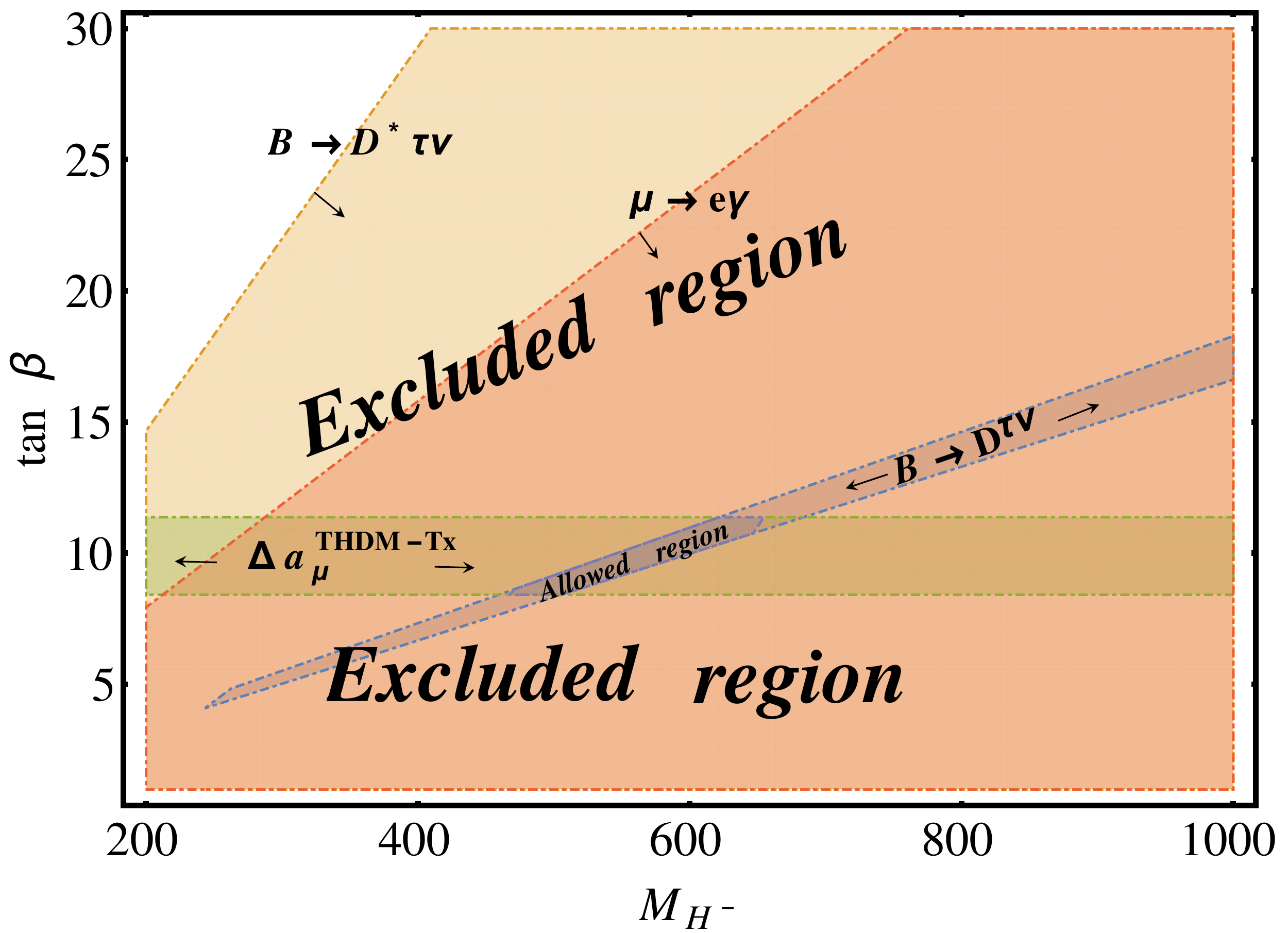}}
\subfigure[ ]{\includegraphics[scale = 0.16]{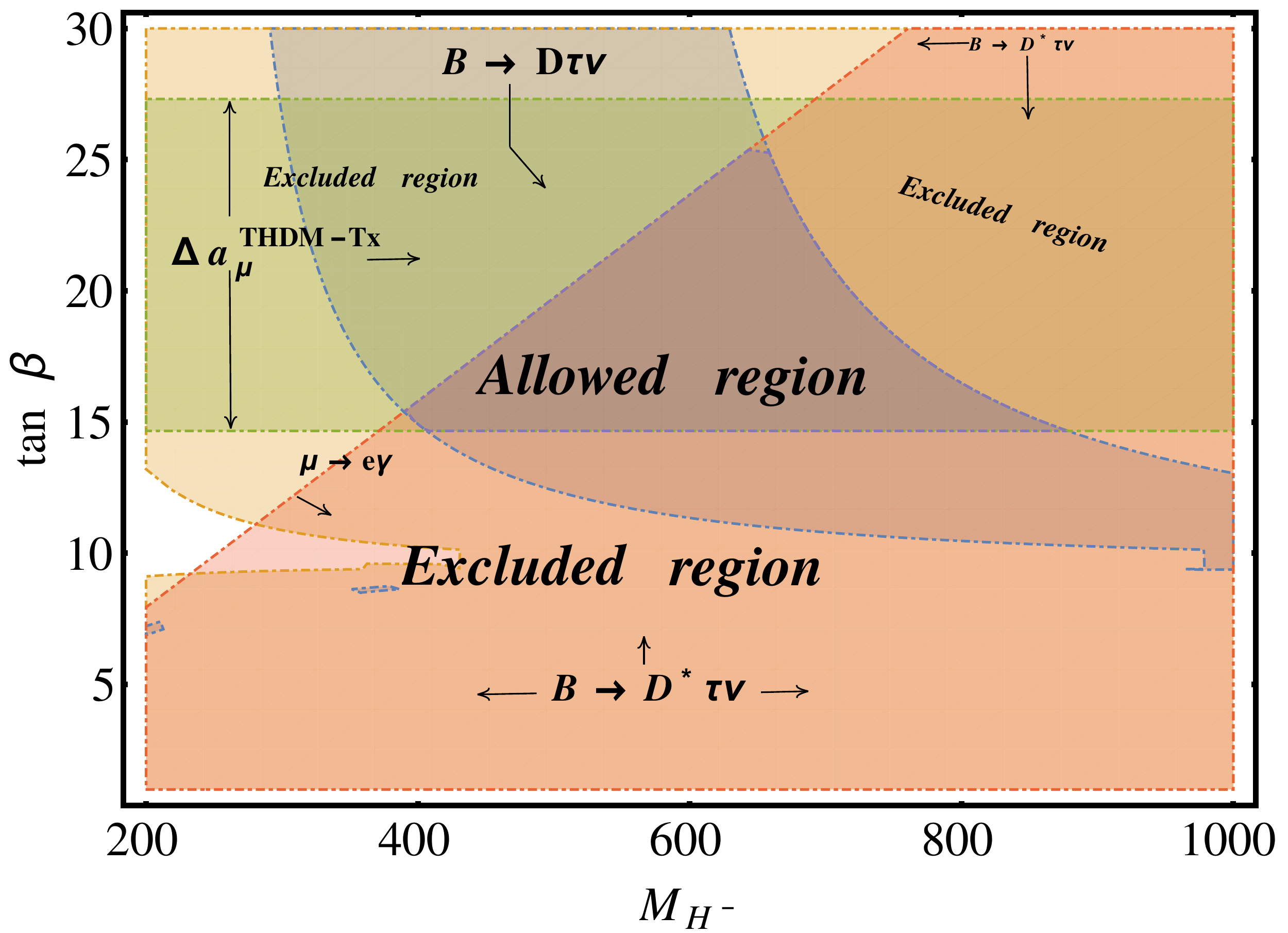}}    
 \caption{Allowed and excluded regions obtained from the processes mediated by charged Higgs bosons for Semi-Parallel (a) and Parallel (b) cases. \label{cargados-semi}}
	\end{figure}

We will now constrain the parameter space for the charged Higgs bosons, following a similar methodology to the one used in the previous section, 
except that we now consider processes mediated by charged scalars: $B\to D(D^*)\tau\nu$, $a_{\mu}^{THDM-Tx}$ and $\mu\to e\gamma$.
The allowed regions for Semi-Parallel and Parallel cases are presented in Figures \ref{cargados-semi}, \ref{cargados-Paralela}. We have performed a thorough analysis of the Complementary case and we found that it was able to satisfy $\mu\to e \gamma$, $B\to D^*\tau \nu$ and $\Delta a_\mu$, but we where unable to find a region that satisfied $B\to D\tau \nu$ and the previous processes. The Complementary case is only able to meet with the limits of neutral Higgs bosons.

		 We find that the most restrictive process  is $B\to D\tau\nu$. We have used  $\gamma_u=0.13$, $\gamma_d=0.1$ and $\gamma_l=1$, $(\alpha-\beta)=\frac{\pi}{2}$, $M_{H^0}=500$ GeV, $M_{A^0}=300$ GeV, $M_{h^0}=125$ GeV $\alpha_1=\alpha_2=0$, $a_1^u=\mathcal{O}(10^{-1})$, $a_1^d=\mathcal{O}(10^{-3})$ , $a_1^l=\mathcal{O}(10^{-3})$.

\section{The LHC Signals}\label{sec:LHC-signal}

In the previous section we determined the allowed parameter space according to current low energy constraints. 
Now, we will use collider constraints. These need to satisfy the constraint of the SM-like Higgs signal observed at the LHC 
by including several production and decay Higgs channels. We shall consider only the production of Higgs bosons by gluon fusion and the decays $h^0\to ZZ^{*}$, $\gamma\gamma$, $\tau^{-} \tau^{+}$, $b\bar{b}$. Then, in order to reproduce  the signal rate for the  SM-like Higgs signals with $m_{h}\simeq 125$ GeV, 
we consider the following ratios:

\begin{equation}
 R_{XX} = \frac{ \sigma( gg\to h^0 ) }{ \sigma( gg\to h_{SM} ) }  
                \frac{ Br(h^0 \to XX) }{ Br (h_{SM} \to XX) }, 
\end{equation}
for $X=\gamma, Z, \tau, b$.

Within the so-called narrow-width approximation, we can write the expression for $R_{XX}$ as follows:

\begin{equation}
 R_{XX} = \frac{ \Gamma(h^0 \to gg) }{ \Gamma(h_{SM} \to gg ) } \, 
                \frac{ Br(h^0 \to XX)}{ Br(h_{SM} \to XX)}.
\end{equation}
The Higgs signals $h^0\to b\bar b$ and $h^0\to \tau^{-}\tau^{+}$ channels have been tested at the LHC due to the relatively large Yukawa couplings. 
In THDM's, the bottom and tau Yukawa coupling are expected to be different from those of the SM. The values of the R-parameters according to current LHC Higgs data such as $R_{Z Z}=1.15^{+0.27}_{-0.23}$, $R_{\gamma \gamma}=1.17^{+0.19}_{-0.17}$,  $R_{b \bar{b}}=0.85 \pm 0.29$ and $R_{\tau^{+} \tau^{-}}=0.79\pm0.26$ can be found in \cite{PDG:2016}. Our evaluation of the values of $R_{XX}$ is done with $(\alpha-\beta)\sim\pi/2$, $\gamma_f=1$, $\alpha_1$=$\alpha_2$=0.

\begin{figure}[H]
\centering
\includegraphics[scale = 0.2,angle=0]{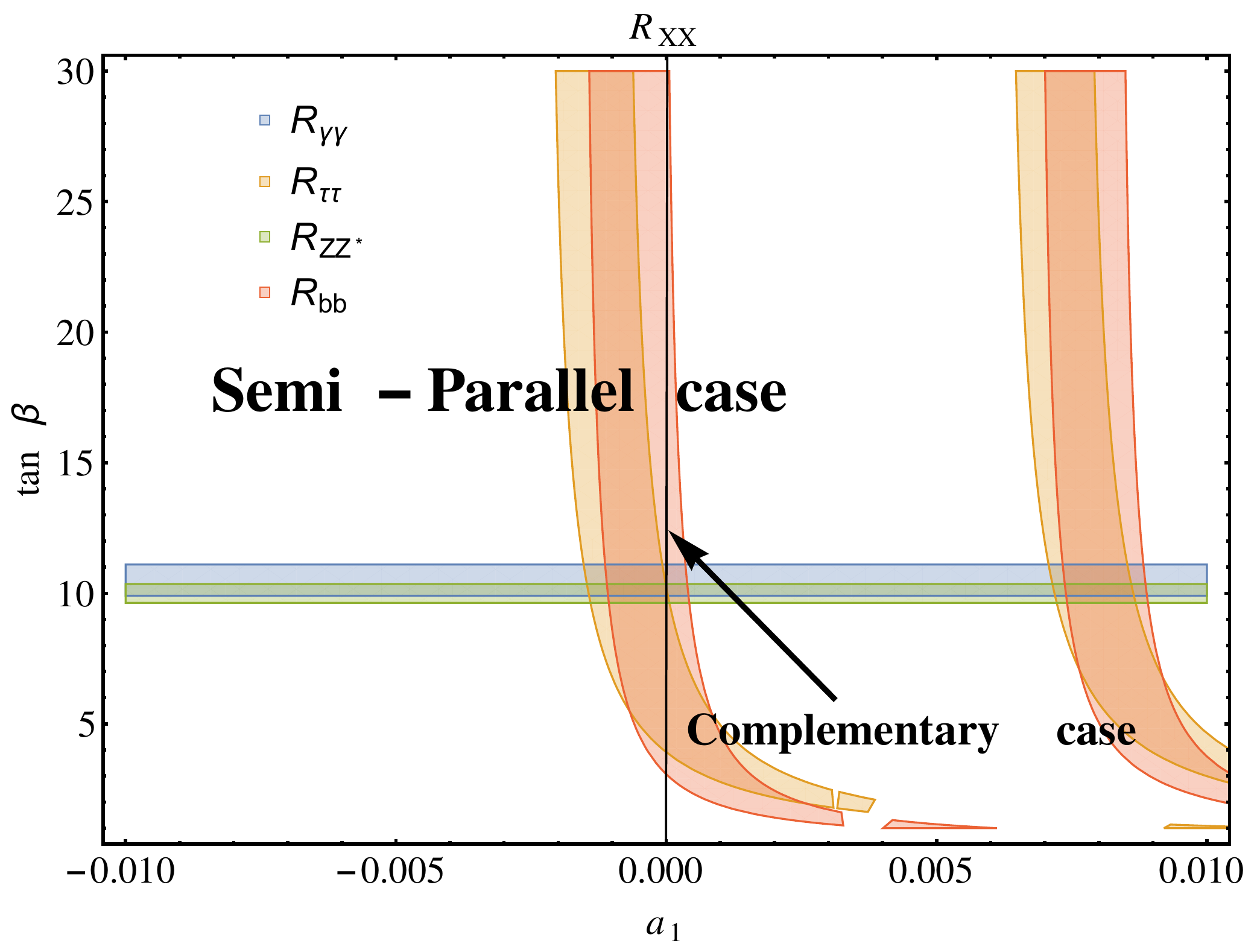}
\caption{Signal strength for $R_{ZZ^{*}}$, $R_{\gamma\gamma}$, $R_{b\bar b}$ and $R_{\tau^{-}\tau^{+}}$. \label{fig:RCa1}}
\end{figure}

The regions that satisfy the constraint for each of the considered channels are presented in Fig. \ref{fig:RCa1}, 
we find several regions in the $\tan\beta-m_H$ plane that satisfy the current signal strength measurements.

\section{Predictions for $h^0 \to \tau \mu,\; t \to c h^0 $}\label{sec:predictions}

The decay $h^0 \to \tau \mu$ provides an interesting signal to probe FV Higgs couplings, 
it was initially studied in \cite{Pilaftsis:1992st, DiazCruz:1999xe}. And subsequent studies on 
detectability of the signal appeared in \cite{phenohlfv}, 
while improved calculations within SUSY and other models appeared in \cite{improvehlfv};
more recent discussions of LFV Higgs decays are presented  in \cite{susyhlfv}.

Another interesting signal to probe FV Higgs couplings are rare top decays, particularly $t \to c h^0 $
which has been studied within the THDM in \cite{Eilam:1990zc, Mele:1998ag,  Bejar:2001sj, Mele:1999zx}, 
 while the SUSY case was considered in \cite{DiazCruz:2001gf, ourtopfcnc}. The search for this mode at LHC was
considered in \cite{AguilarSaavedra:2000aj}.

\subsection{$h^0\to\tau\mu$ decay}
The first search for the Lepton Flavor Violating (LFV) decay $h^0 \to \tau \mu$ was performed by CMS \cite{limitehtaumu} and ATLAS \cite{Aad:2016blu},
 at a center of mass energy of $\sqrt{8}$ TeV with an integrated luminosity of 19.7$fb^{-1}$. They reported a slight signal excess 
 with a significance of 2.4$\sigma$, and give a limit on BR($h^0\to\tau\mu$)$<1.51\%$ at 95$\%$ confidence level.

\begin{figure}[H]
\centering
      {\includegraphics[scale = 0.5]{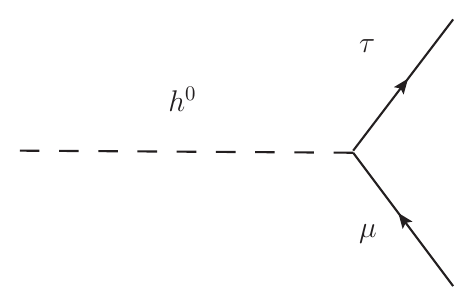}}
\caption{Feynman diagram for $h^0 \to \tau \mu$ decay.\label{fd:htaumu}}
\end{figure}

\begin{figure}[H]
\centering
    \subfigure[ ]{\includegraphics[scale = 0.225]{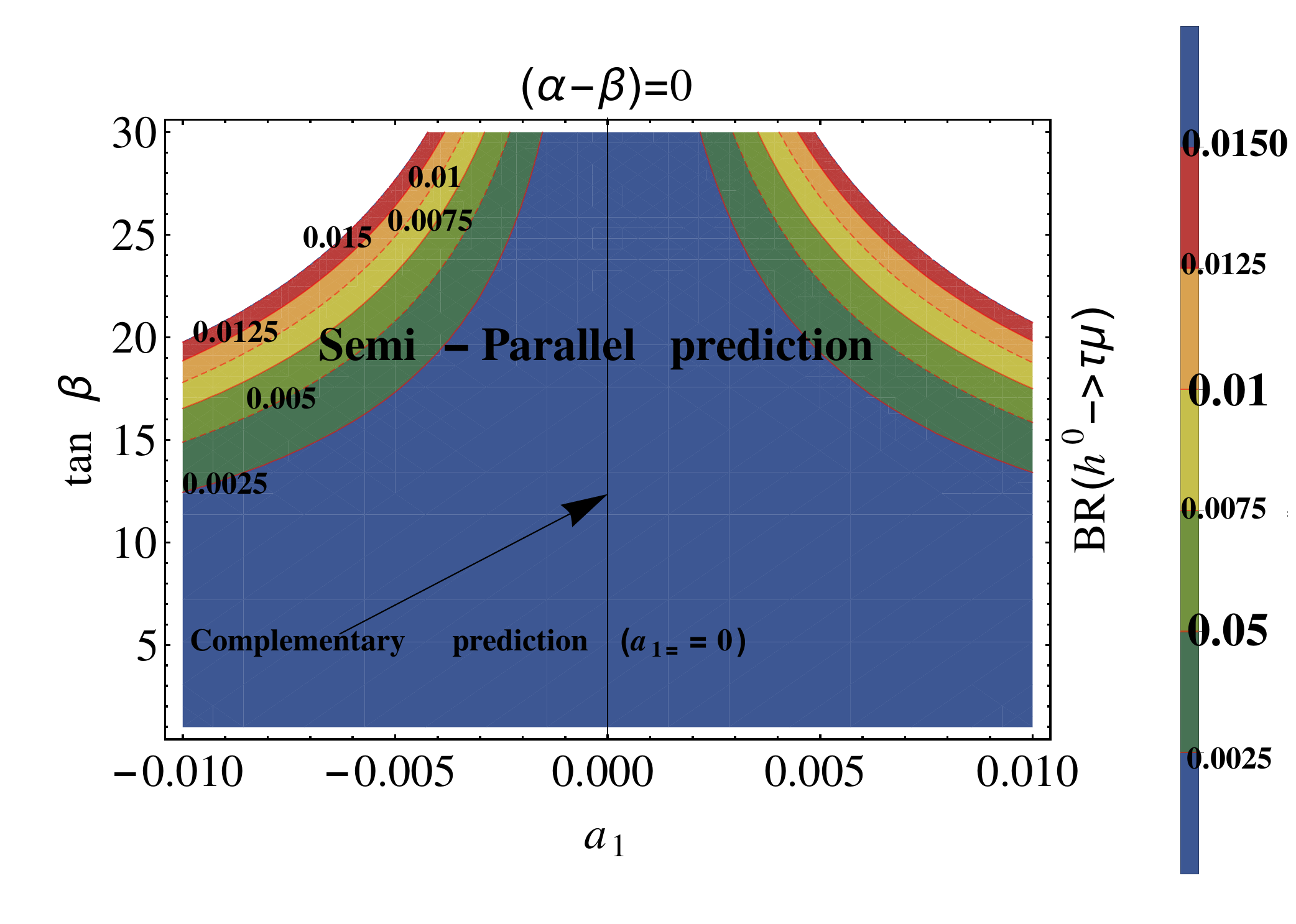}\label{htaumucero}} 
    \subfigure[ ]{\includegraphics[scale = 0.225]{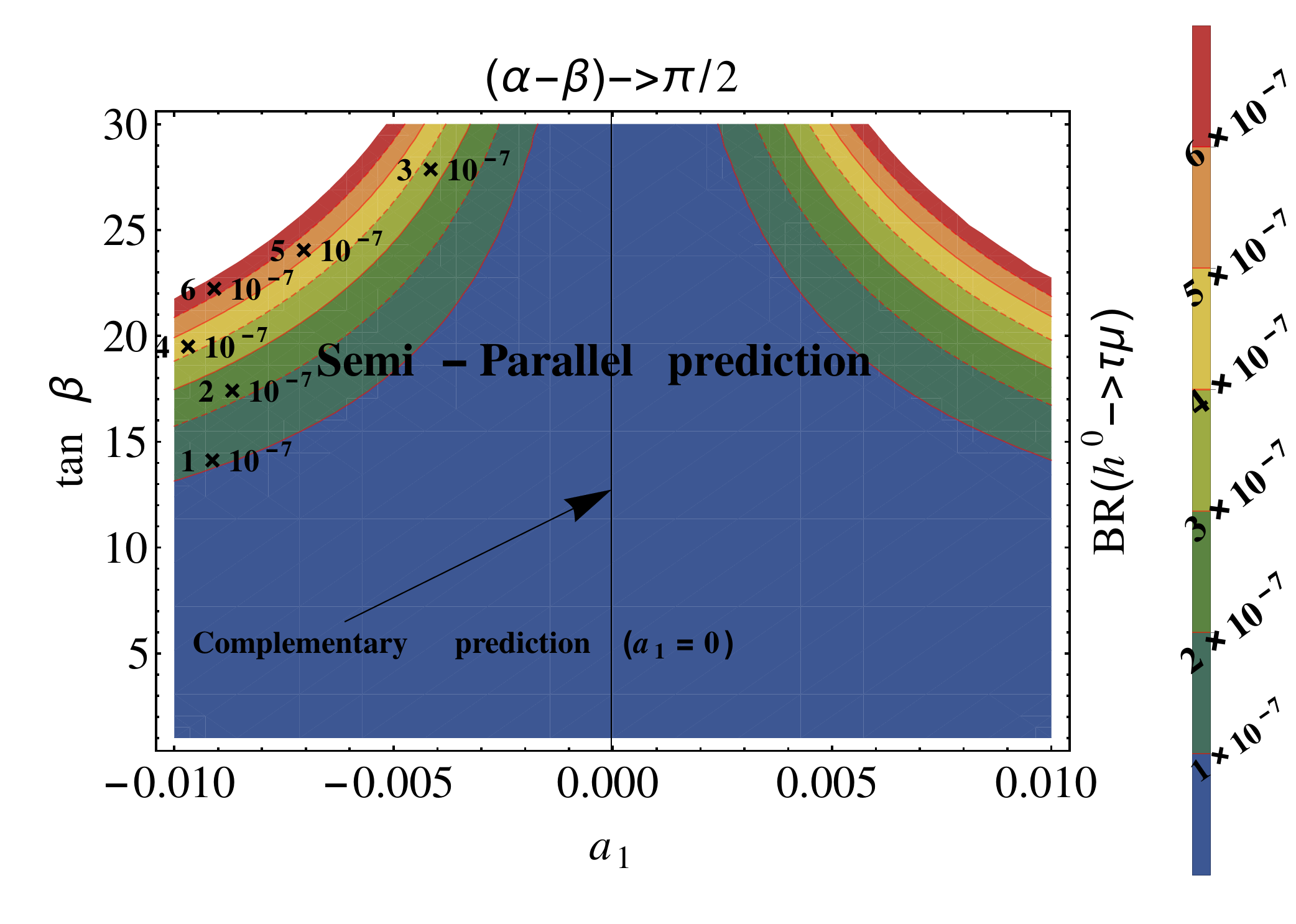}\label{htaumupi2}} 
   \caption{Branching ratios of the LFV decay $h^0 \to \tau\mu$ as a function of $\tan{\beta}$ and the parameter $a_1$. The vertical line, 
   whose equation is $a_1=0$, corresponds to the prediction of the complementary case, while the prediction for the Semi-Parallel case is 
   presented in the colored points in the $t_{\beta}$-$a_1$ plane.\label{htaumu}}
	\end{figure}

The Feynman diagram at tree level of the $h^0 \to \tau \mu$ decay is shown in Fig. \ref{fd:htaumu} 

\begin{equation}
\Gamma(h^0 \to \tau \mu) = \frac{ g^2 m_\mu m_\tau}{32 \pi M_W^2} \Big|\eta_{\tau\mu}^{h^0}\Big|^2 \beta^3 M_{h^0},
\end{equation}
where $\beta= \Big(1 - {(r_{\tau h^0} + r_{\mu h^0})^2}\Big) \sqrt{{(r_{\tau h^0}^2 - r_{\mu h^0}^2 - 1)^2} - 4{r_{\mu h^0}^2}}$, $r_{\ell h^0}=m_{\ell}/M_{h^0}$, $M_{h^0}$, $m_{\tau}$, $m_{\mu}$ and $\eta_{\tau\mu}^{h^0}$ are the SM-like Higgs boson, tau and muon masses and $h^0\tau\mu$ coupling, respectively.
As mentioned previously we have been considering $(\alpha-\beta)=\pi/2$ but from eq.\ref{lagrangiano} we identify that this coupling is zero. 
That is why, we take the limit $(\alpha-\beta)\to\pi/2$. We also consider the scenarios $(\alpha-\beta)=0$, which corresponds to the case where 
there are no flavor changing neutral currents via a heavy scalar, and $(\alpha-\beta)=\pi/3$ which corresponds to an intermediate scenario. 
In Fig. \ref{htaumu} we show the branching ratios of the decay $h^0\to\tau\mu$ for these scenarios as a function of $\tan\beta$ and the 
parameter $a_1$ for the complementary case $(a_1=0)$ and Semi-Parallel case $(-0.01\leq a_1\leq 0.01)$. 
The values $-0.01\leq a_1 \leq 0.01$ are used because this interval satisfies the experimental bounds and values of the process that we 
considered to restrict the parameter space of our model. Fig. \ref{htaumucero} corresponds to the $(\alpha- \beta)=0$ scenario and Fig. \ref{htaumupi2} 
corresponds to $(\alpha- \beta)\to\pi/2$ scenario. We find that this scenario gives the smallest contributions with a $BR\sim 10^{-7}$.  
We only present the branching ratios that fall within the current limit BR($h^0\to\tau\mu$)$< 1.51\%$  \cite{limitehtaumu}. 	
The graph corresponding to the scenario $(\alpha-\beta)=\pi/3$ is shown in appendix \ref{Graficas}.

%
		 \subsection{$t\to ch^0$ decay}
\begin{figure}[H]
\centering
   {\includegraphics[scale = 0.5]{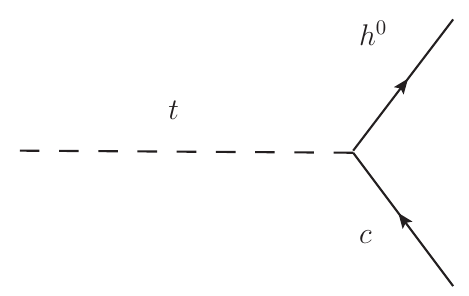}}
\caption{Feynman diagram for $t \to ch^0$.\label{fd:tophc}}
\end{figure}

\begin{figure}[H]
\centering
 \subfigure[ ]{\includegraphics[scale = 0.225]{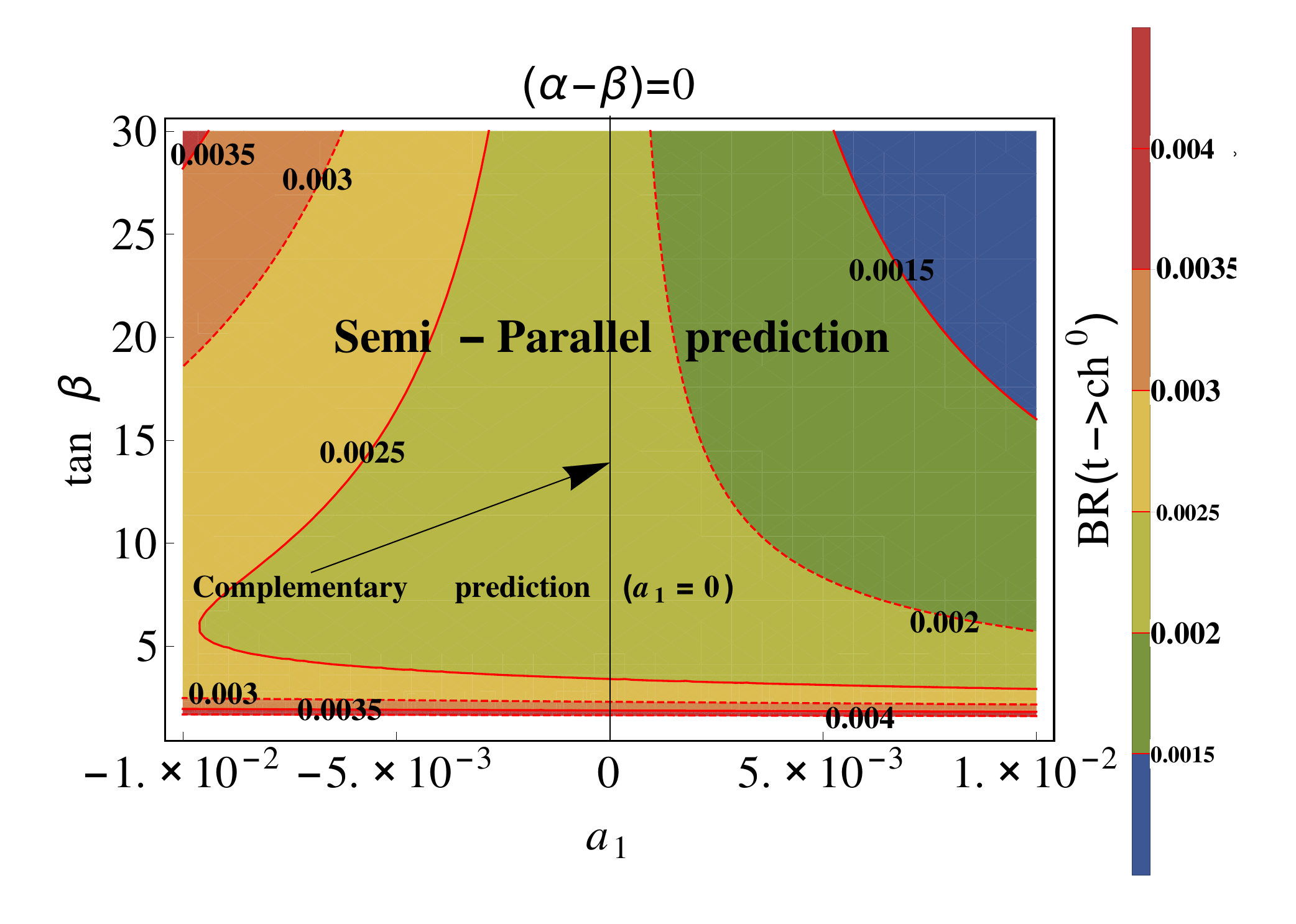}\label{tchcero} }  
 \subfigure[ ]{\includegraphics[scale = 0.225]{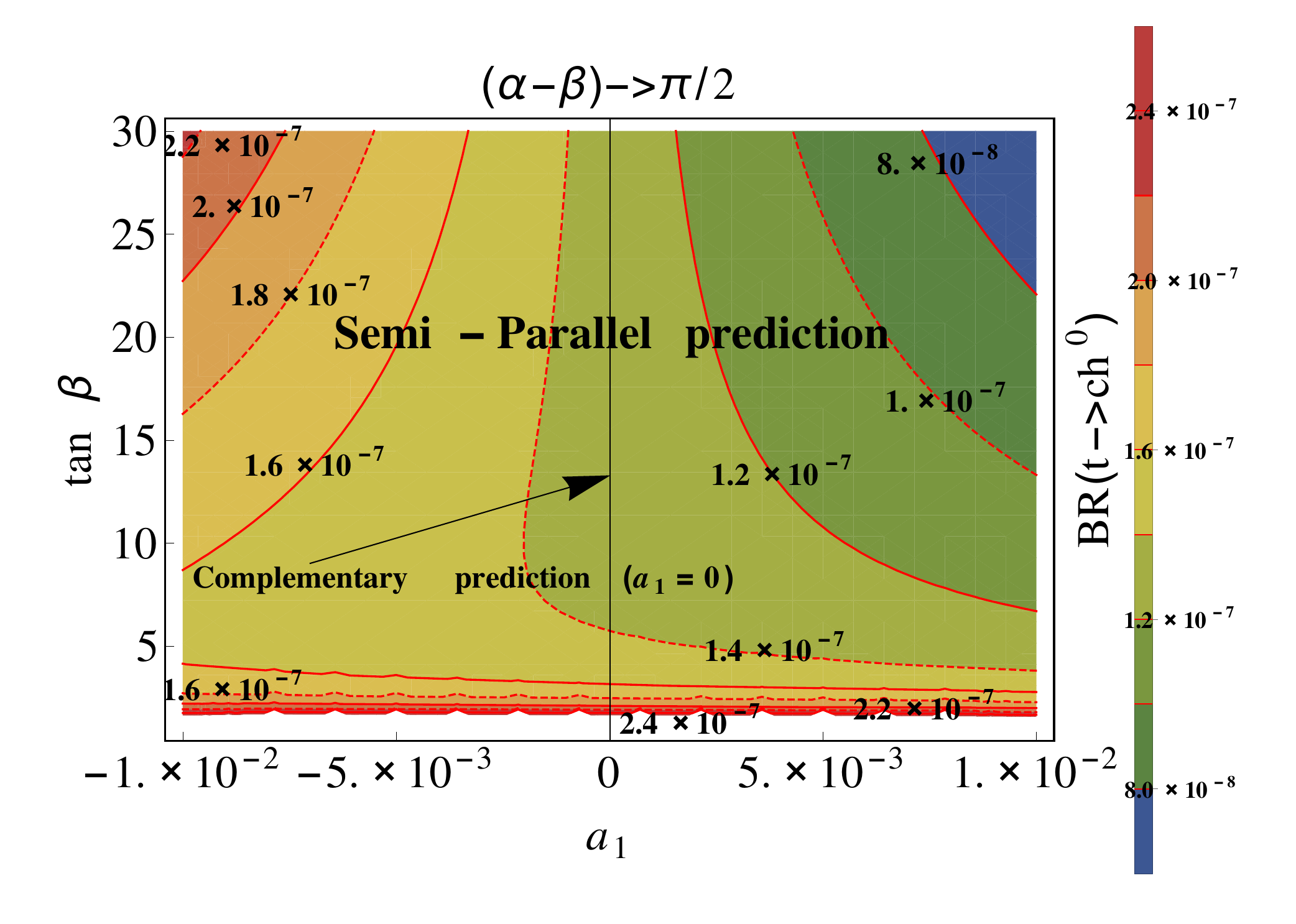}\label{tchpi2}}    
      \caption{Branching ratios of the rare decay $t\to ch^0$ as a function of $t_{\beta}$ and the parameter $a_1$ for (a) $(\alpha-\beta)=0$ and (b) $(\alpha-\beta)\to\pi/2$. The vertical line, whose equation is $a_1=0$, corresponds to the prediction of the complementary case, while the prediction for the Semi-Parallel case is presented in the colored points in the $\tan{\beta}$-$a_1$ plane.\label{tch}}
\end{figure}

The Feynman diagram of the $t \to c h^0$ decay is shown in Fig. \ref{fd:tophc},
and the decay width is given by
\be
\Gamma(t\to ch) = \frac{ m_t}{16 \pi}\Big|\eta_{tc}^{h^0}\Big|^2 \Big[(1 + r_c)^2 - r_{ht}^2\Big]  
\times \sqrt{1 - (r_{ht} + r_{ch})^2} \sqrt{1 - (r_{ht} - r_{hc})^2}.
\ee
 In Fig. \ref{tch} we show the Branching ratio of the decay $t\to ch^0$ as a function of $\tan\beta$ for the complementary case $a_1=0$ and Semi-Parallel case; the values $-0.01\leq a_1 \leq 0.01$ are used because this interval satisfies the experimental bounds and values of the process that we considered to restrict the parameter space of our model. Fig. \ref{tchcero} corresponds to the $(\alpha-\beta)=0$ scenario and Fig. \ref{tchpi2} to $(\alpha- \beta)\to\pi/2$ scenario. We find that this scenario gives the smallest contributions with a $BR\sim 10^{-7}$.  We only present the branching ratios that fall within the current limit BR($t\to ch^0$)$< 5.6\times 10^{-6}$
The graph corresponding to the scenario $(\alpha-\beta)=\pi/3$ is shown in appendix \ref{Graficas}.


\section{Conclusions}\label{sec:conclusio}
We have considered the possibility that the Yukawa matrices can be constructed in such a way that they produce a specific 
hermitian mass matrix, which we classify as Semi-Parallel and Complementary textures. This assumption of different matrix 
textures is relevant in the study of LFV decays and rare quark decays, because the flavor violating couplings for each of these 
new signals depends on the Yukawa elements $Y_{ij}$. 
Through current experimental bounds from low-energy processes: $\Delta a_\mu$, $K-\bar{K}$ and $B-\bar{B}$ mixing, 
$B\rightarrow D(D^*)\tau\nu$, $B_s^0 \to \mu^+\mu^-$, $l_i\rightarrow l_j \gamma$ and $\tau^-\rightarrow \mu^-\mu^+\mu^-$, 
we restrict our parameter space and find that small values of  $\tan\beta$ and $450\lesssim M_H\leq1000$ GeV are favored. 
We then incorporate the LHC Higgs signal strengths $R_{\gamma\gamma}$, $R_{ZZ^*}$, $R_{b\bar{b}}$ and $R_{\tau^{-}\tau^{+}}$. 
And finally, using all the aforementioned restriction we calculate the branching ratios of the LFV Higgs decay ($h^0\to\tau\mu$), as well as the rare top decay $t\to ch^0$, 
and we find that the highest contributions are of the order $\mathcal{O}(10^{-2})$ and $\mathcal{O}(10^{-3})$, for each one. 
Therefore, our analysis seems to suggests that the case that best satisfies the current experimental bounds and measurements is the 
Semi-parallel case, while the complementary case offers the simplest pattern of Flavor violating Higgs couplings, as they depend only on a few parameters.

\begin{acknowledgments}
We acknowledge support from CONACYT-SNI (Mexico). 
\end{acknowledgments}


  \clearpage

 \newpage\appendix


\section{Yukawa matrices: Cases 2-6 .}\label{Txt-cases2-6}

\begin{center}
Case 2:
\end{center}

        \begin{displaymath}
        {Y}_1 =  \left( \begin{array}{ccc}
0 & d & 0 \\
d^* & c & 0 \\
0 & 0 & 0
        \end{array} \right),
{Y}_2 =  \left(
        \begin{array}{ccc}
0 & 0 & 0 \\
0 & 0 & b \\
0 & b^* & a
        \end{array} \right).
                \end{displaymath}
\begin{center}
Case 3:
\end{center}
        \begin{displaymath}
        {Y}_1 =  \left( \begin{array}{ccc}
0 & d & 0 \\
d^* & 0 & b \\
0 & b^* & 0
        \end{array} \right),
{Y}_2 =  \left( \begin{array}{ccc}
0 & 0 & 0 \\
0 & c & 0 \\
0 & 0 & a
        \end{array} \right).
                \end{displaymath}
\begin{center}
Case 4:
\end{center}
\begin{displaymath}
{Y}_1 =  \left( \begin{array}{ccc}
0 & 0 & 0 \\
0 & c & b \\
0 & b^* & 0
        \end{array} \right),
{Y}_2 =  \left( \begin{array}{ccc}
0 & d & 0 \\
d^* & 0 & 0 \\
0 & 0 & a
        \end{array} \right).
                \end{displaymath}
\begin{center}
Case 5:
\end{center}
\begin{displaymath}
{Y}_1 =  \left( \begin{array}{ccc}
0 & 0 & 0 \\
0 & 0 & b\\
0 & b^* & 0
        \end{array} \right),
{Y}_2 =  \left( \begin{array}{ccc}
0 & d & 0 \\
d^* & c & 0 \\
0 & 0 & a
        \end{array} \right).
                \end{displaymath}
\begin{center}
Case 6:
\end{center}
\begin{displaymath}
{Y}_1 =  \left( \begin{array}{ccc}
0 & d & 0 \\
d^* & 0 & 0 \\
0 & 0 & 0
        \end{array} \right),
{Y}_2 =  \left( \begin{array}{ccc}
0 & 0 & 0 \\
0 & c & b \\
0 & b^* & a
        \end{array} \right).
                \end{displaymath}
\begin{center}
Case 7:
\end{center}
\begin{displaymath}
{Y}_1 =  \left( \begin{array}{ccc}
0 & 0 & 0 \\
0 & c & 0 \\
0 & 0 & 0
        \end{array} \right),
{Y}_2 =  \left( \begin{array}{ccc}
0 & d & 0 \\
d^* & 0 & b \\
0 & b^* & a
        \end{array} \right).
                \end{displaymath}

\section{Scenario $(\alpha-\beta)=\pi/3$ .}\label{Graficas}
\begin{figure}[H]
\centering
    \subfigure[ ]{\includegraphics[scale = 0.22]{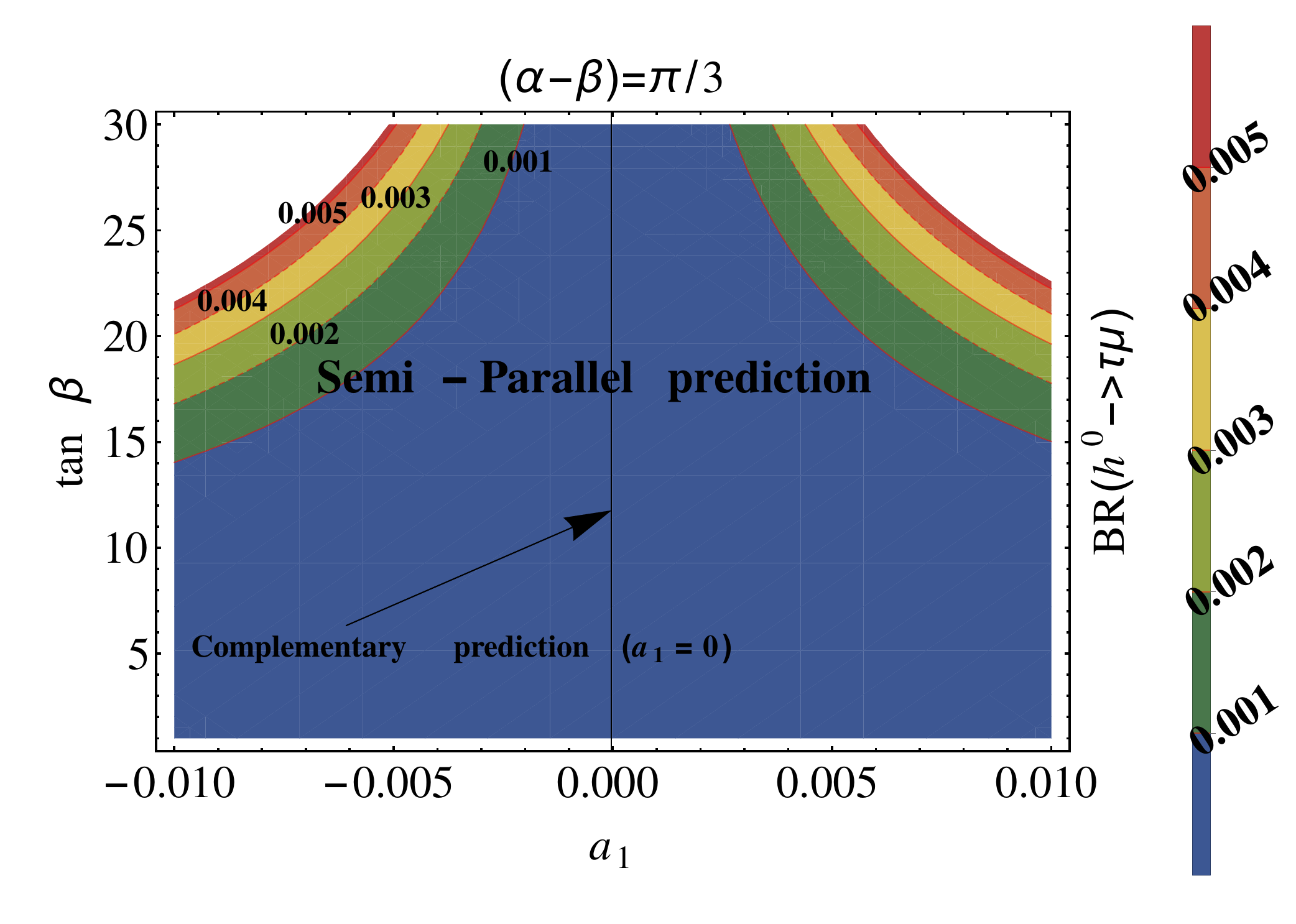}} 
    \subfigure[ ]{\includegraphics[scale = 0.22]{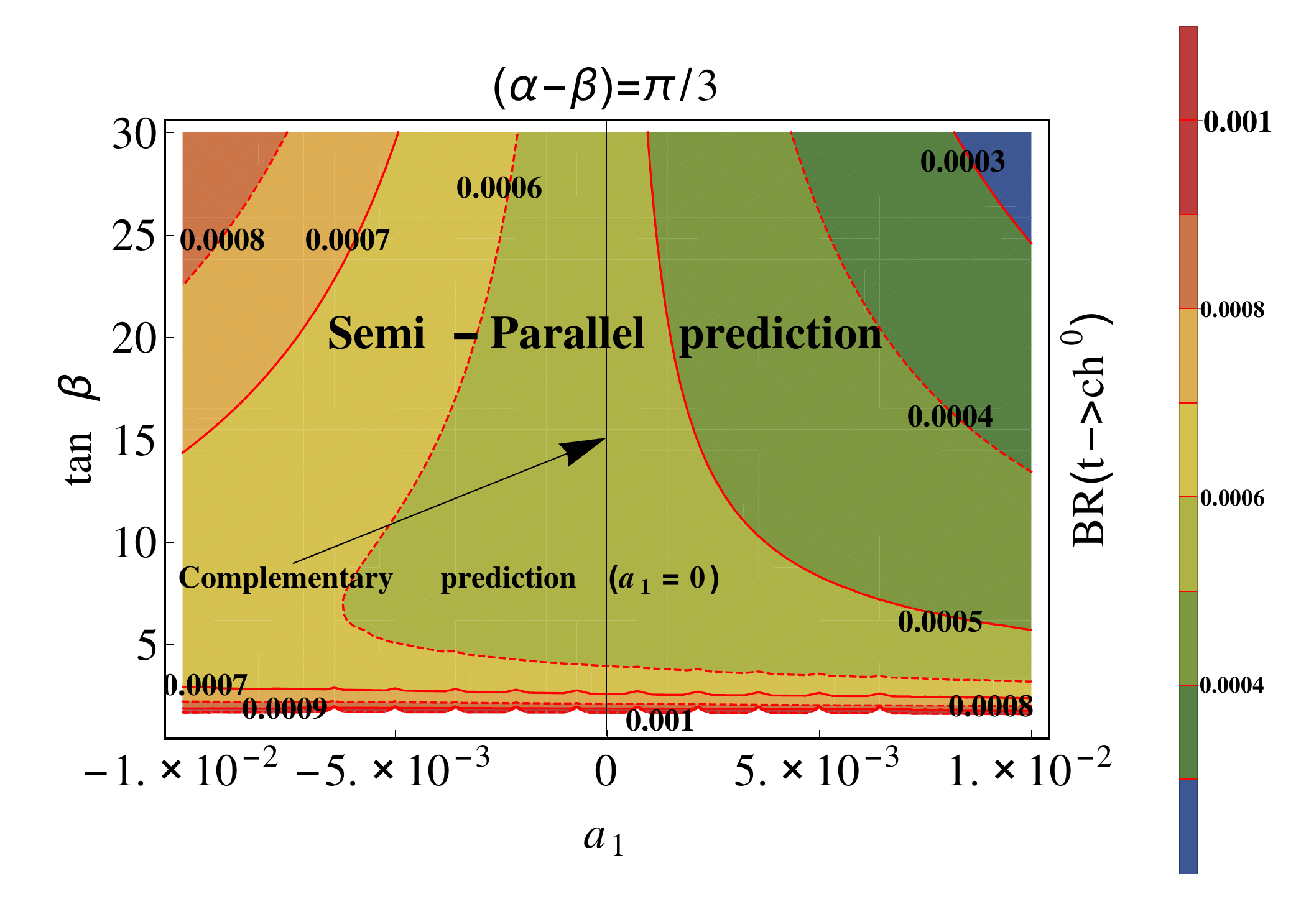}}   
   \caption{(a) Branching ratios of the LFV decay $h^0 \to \tau\mu$ as a function of $t_{\beta}$ for the scenario $(\alpha-\beta)=\frac{\pi}{3}$. (b) Branching ratios of the rare top decay $t \to ch^0$ for the scenario $(\alpha-\beta)=\frac{\pi}{3}$.}
	\end{figure}
	




%



\end{document}